\documentstyle[epsf,amssymb]{article}

\renewcommand{\theequation}{\arabic{section}.\arabic{equation}}

\begin{document}
\begin{center}
{\Large {\bf Phase Structure of a 3D Nonlocal U(1) Gauge Theory:
Deconfinement by Gapless Matter Fields}}

\vskip 1cm

{\Large Gaku Arakawa and Ikuo Ichinose}  \\
{ Department of Applied Physics,
Nagoya Institute of Technology, Nagoya, 466-8555 Japan}
\vskip 0.5cm
{\Large Tetsuo Matsui} \\
{Department of Physics, Kinki University, 
Higashi-Osaka, 577-8502 Japan } 
\vskip 0.5cm
{\Large Kazuhiko Sakakibara} \\
{Department of Physics, Nara National College of Technology, 
Yamatokohriyama, 639-1080 Japan}
\vskip 0.5cm
{\Large Shunsuke Takashima}  \\
{ Department of Applied Physics,
Nagoya Institute of Technology, Nagoya, 466-8555 Japan}

\end{center}

\begin{center} 
\begin{bf}
Abstract
\end{bf}
\end{center}
In this paper, we study a 3D compact U(1) lattice gauge theory
with a variety of nonlocal interactions that simulates 
the effects of gapless/gapful matter fields.
We restrict the nonlocal interactions among gauge variables only to those
along the temporal direction and adjust their coupling constants optimally 
to simulate the isotropic nonlocal couplings of the original model.
This theory 
is quite important to investigate the phase structures of
QED$_3$ and
strongly-correlated electron systems like the 2D quantum spin 
models, the fractional quantum Hall effect, the t-J model of 
high-temperature superconductivity.
We perform numerical studies of this theory to find 
that, for a certain class of power-decaying couplings, 
there appears a second-order
phase transition to the deconfinement phase as the gauge coupling 
constant is decreased.
On the other hand, for the exponentially-decaying coupling,
 there are no signals for second-order phase transition.
These results indicate the possibility that introduction of
sufficient number of massless matter fields destabilizes  the permanent 
confinement in the 3D compact U(1) pure gauge theory due to instantons.

\newpage
\section{Introduction}
Gauge theories and their associated concepts 
have played important roles not only in  elementary
particle physics but also in condensed matter physics.
For example, the conventional superconducting phase transition
is characterized as a change of the gauge dynamics of
the electromagnetic U(1) gauge symmetry.
Also, for a variety of strongly-correlated electron systems, 
it has been recognized that their phase structures and the properties 
of low-energy excitations  
are naturally described by using the terminology of 
gauge theories\cite{im1}.
As such low-energy ``quasi-particles", 
the composite fermions/bosons in the fractional quantum Hall 
states\cite{FQHE} and 
the holons and spinons in the t-J model of high-$T_c$ cuprates\cite{CSS}
have been proposed. 
It was argued\cite{pfs} that their unconventional properties like 
fractionality and their existence itself may be explained by 
a confinement-deconfinement phenomenon of the
gauge dynamics of the effective gauge theories derived from the original models.
This interesting idea is still controversial\cite{nayak}, but
certainly warrants further investigation.

Most of the problems in gauge-theoretical studies 
on the strongly-correlated electron systems reduces to studying the
phase structures of gauge theories coupled to {\it gapless}
relativistic/nonrelativistic matter fields.
In the elementary particle physics, it is generally believed that
the phase structures of gauge systems coupled to  matter
fields are {\em  not} easy to study analytically because
introduction of matter fields results in the lack
of simple order parameters such as Wilson loops for 
pure gauge systems\cite{FS}.
Furthermore, inclusion of gapless fermions make numerical simulations 
a difficult task since one must face quite nonlocal interactions 
generated by integrating over fermion variables.
Sometimes couplings to gapless matter 
fields change the universality class of the gauge system under 
consideration from that of the pure gauge model.
A good example is the four-dimensional (4D) QCD coupled with 
light quarks in which the number of light quarks strongly 
influences its phase structure\cite{QCD}.

In this paper, we address this problem of gauge dynamics of coupled 
systems. 
Specifically, we are interested in the U(1) lattice gauge 
theory (LGT) with gapless/gapful and relativistic/nonrelativistic 
matter fields in 
three dimensions (two spatial dimensions at zero temperature). 
This theory of course covers the important model QED$_3$\cite{QED}.
It appears also as a main part of the effective gauge theory of
the strongly-correlated electron systems mentioned above, and so
plays an important role in studying these systems\cite{pfs,AFH}.
In the ordinary 3D compact U(1) pure gauge system 
(i.e., without matter fields) with local interactions, it is 
established that only the confinement phase is realized because of 
instanton condensation\cite{polyakov}.
For the case with additional massless (gapless) matter fields 
coupled to the U(1) gauge 
field, recent studies give controversial results on the possibility
of a deconfinement phase; it is supported in
Ref.\cite{pfs,QED3} whereas it is denied in Ref.\cite{css-}.

Our approach to this problem is by (i) introducing an effective theory 
of the original theory and (ii) studying its phase structure numerically.
As the  effective theory we use a 3D U(1) pure LGT
with nonlocal interactions among gauge variables.  
These nonlocal interactions are along the temporal direction
and mimic the effect of matter fields.
We consider exponentially-decaying  interactions for  massive
matter fields (i.e., fields with gaps) 
and power-decaying interactions for massless (gapless) fields.
We shall see that certain cases of power-decaying interactions
exhibit second-order phase transitions which separate
the confinement phase and the deconfinement phase\cite{AIMS}.
The existence of the deconfinement phase in the effective theory  
indicates that it is realized in the original model if the number 
of gapless matter fields is sufficiently large.
This result is in agreement with the results of Ref.\cite{pfs,QED3}.

The rest of the paper is organized as follows.
In section 2, we briefly survey the effect of matter fields
in several aspects. 
In section 3, the original model and its effective nonlocal gauge model 
are explained. 
Section 4 is devoted for numerical calculations.
We  calculate the internal energy, specific heat, expectation
values of Polyakov loops, Wilson loops, and density of instantons.
All these quantities indicate a second-order 
confinement-deconfinement phase transition 
(CDPT) for the gauge models with sufficiently long-range correlations
among the gauge variables.
In section 5, we study tractable low-dimensional spin models,
which are obtained by simple reduction of the gauge degrees of freedom
in the nonlocal effective model.
Then we obtain an intuitive picture of the CDPT of the present long-range
U(1) gauge theories.
Section 6 is devoted for conclusion.
In Appendix, the effective nonlocal model 
is studied by the low- and high-temperature expansions.
The analytic expressions of the internal energy and the specific
heat are in good agreement with the numerical calculations
in Sect.4.


\section{Effects of Matter Fields on Gauge Dynamics }

In this section, we briefly review the effects of matter fields
upon the U(1) gauge dynamics in several aspects.

\subsection{Weak-coupling regime in noncompact/compact U(1) gauge theory}

One may generally expect that inclusion of massless matter fields to 
a system of gauge field may drastically change the gauge dynamics 
at long wavelengths.
For the case that {\em relativistic and massless} matter fields 
are coupled to a U(1) gauge field $A_\mu(x)\ (\mu=0,1,2)$ in 3D, 
the one-loop radiative correction of matter fields at weak gauge couplings
generates 
the following nonlocal term in the  effective action of $A_\mu(x)$;
\begin{equation}
\Delta A \propto e^2\int d^3x \int d^3y \sum_{\mu, \nu}F_{\mu\nu}(x)
{1\over |x-y|^2}F_{\mu\nu}(y),\ 
F_{\mu\nu}=\partial_\mu A_\nu-\partial_\nu A_\mu.
\label{LA}
\end{equation}
It is obvious that the above term  strongly suppresses fluctuations
of $A_{\mu}(x)$ at long distances due to the factor $|x-y|^{-2}$. 
The above nonlocal terms are leading at low energies and momenta
{\em if} its effective coupling constant does not vanish at the infrared limit.
Then the potential energy $V(r)$ between two charges separated by 
distance $r$ changes drastically,
\begin{eqnarray}
V(r) \propto \log r \rightarrow V(r) \propto \frac{1}{r}.
\label{potential}
\end{eqnarray}

In the compact U(1) gauge theory, topologically nontrivial excitations
(e.g. instantons) of $A_{\mu}(x)$ appear whose effect at small 
gauge coupling $e$ can 
be estimated by replacing
the field strength $F_{\mu\nu}(x)$ in Eq.(\ref{LA}) by 
$F_{\mu\nu}(x)-2\pi n_{\mu\nu}(x)$,
where $n_{\mu\nu}$ is an integer field whose rotation
measures the instanton number(density) $\rho(x) =
\epsilon_{\mu\nu\lambda}\partial_\mu n_{\nu\lambda}(x)$.
In the gauge model with the usual Maxwell term, the potential energy between
a pair of instantons at distance $r$ is Coulombic, $V_{\rm ins}(r) \sim 1/r$.
However, the long-range action (\ref{LA}) modifies $V_{\rm ins}(r)$
to a long-range one\cite{QED3},
\begin{eqnarray}
V_{\rm ins}(r) \propto \frac{1}{r} \rightarrow V_{\rm ins}(r) \propto \log r.
\end{eqnarray}
Recently, it was argued that a gas of charged particles with the
long-range interaction like $\log r$ exhibits a phase 
transition between a dilute gas of dipoles and a plasma\cite{plasma}.
This result implies that instantons in the long-range gauge theories 
form dipole pairs and do not condense in the weak-coupling region.
Since the confinement phase of gauge dynamics requires a condensation
of isolated instantons (a plasma phase), this result leads to a 
deconfinement phase of gauge dynamics there.\\

\subsection{The 3D CP$^{N-1}$ model on the criticality}

As an example of the nonperturbative effect of matter fields,
let us consider   the CP$^{N-1}$ model in a 
3D continuum.
The action of the model is given as 
\begin{equation}
A_{\rm CP}=
\int d^3x\Big[\frac{1}{f}|(\partial_\mu+iA_\mu) z|^2+
\sigma\Big(|z|^2-1\Big)\Big],
\label{CP}
\end{equation}
where $z_a(x)\; (a=1,2,\cdots,N)$ is the CP$^{N-1}$ field
satisfying the local constraint,
$\sum_{a=1}^N |z_a(x)|^2 =1$ for each $x$
via the  Lagrange multiplier field $\sigma(x)$, and
$A_\mu(x)$ is the auxiliary U(1) gauge field.

At large $N$, the $1/N$ expansion is reliable and predicts
 a second-order phase transition at the  critical coupling 
 $f=f_c$ \cite{CPN}.
For $f>f_c$, the model is in the disordered-confinement phase 
in which  $\langle \sigma(x) \rangle \neq 0$,
whereas for $f<f_c$, the model is in the ordered-Higgs phase
in which $\langle \sigma(x) \rangle = 0$ and  
the CP$^{N-1}$ field has a nonvanishing expectation value
like $\langle z_N(x)\rangle =v_0 \neq 0$.
As a result, the gauge field $A_\mu$ acquires a finite mass ($\propto v_0$) 
by the Anderson-Higgs mechanism, and 
the low-energy excitations are gapless $z_a(x)\; (a\neq N)$ fields.

Recently, considerable interests have been paid on the question how the gauge 
field behaves just at the {\em critical point} $f=f_c$\cite{yoshioka}.
In the leading order of the $1/N$ expansion, 
it is shown that all the components $z_a(x)\; (a=1,2,\cdots, N)$
are  massless and the gauge field $A_\mu$ acquires the nonlocal
``kinetic term" of Eq.(\ref{LA}).
This implies that, {\em at the critical point}, the nonperturbative
fluctuations of $A_\mu$ like instantons are suppressed, so 
the gauge dynamics at $f=f_c$ is in the {\em Coulomb phase} 
with the potential $V(r) \sim 1/r$ as Eq.(\ref{potential}).

Then we have numerically studied the  3D CP$^1+$ U(1) 
LGT from the above point of view\cite{TIM}.
The numerical results of the CP$^1$ model show a 
second-order transition and the 
confinement phase is realized at the critical point.
However, calculations of CP$^{N-1}$ model with $N=3,4,5$ show 
that 
the topologically nontrivial configurations are suppressed more 
as $N$ increases. We expect 
that there exists a critical value of $N = N_c$, and
the  {\em deconfinement phase } is realized at the critical point for
$N_c < N$ in compatible with the large-$N$ analysis.
We note that a similar phenomenon of inducing a deconfinement phase
by a plenty of massless matter fields has been established in lattice 
QCD in 4D.
For a sufficiently large number of light flavors $N_f>7$, the model 
stays in the
deconfinement phase even the pure gauge term is missing\cite{QCD}.

\subsection{Nonrelativistic fermions in strongly-correlated electron systems}

Nonrelativistic fermions  are distinguished from relativistic 
fermions by the properties;
(i) they propagate only in the positive direction of the imaginary
time in path-integral formulation, and (ii) they form a Fermi surface
(line). 
In strongly-correlated electron systems, one faces nonrelativistic 
fermions not only in the original models but also in their
effective gauge models. 
In studying the fractionalization phenomena of electrons  like
the charge-spin separation (CSS) \cite{CSS}
in high-temperature superconductivity 
and the particle-flux separation (PFS) 
in fractional quantum Hall systems \cite{FQHE}, we
regard an electron $C_x$ (we suppress the spin index for simplicity) 
at the site $x$ of a 2D spatial lattice as a composite of
a fermion $A_x$ and a boson $B_x$ \cite{pfs} as
\begin{eqnarray}
C_x &=& A_x^\dagger B_x.
\end{eqnarray}
To assure the correct physical space composed of $C_x$,
 one  imposes the local constraint for the physical states  
 $|\rm{phys}\rangle$ as
\begin{eqnarray}
\big(A_x^\dagger A_x + B_x^\dagger B_x -1\big)|{\rm phys} \rangle =0\
{\rm for\ each\ } x.
\label{constraint}
\end{eqnarray}
The hopping term of electrons 
(e.g., the $t$-term of the t-J model) may be rewritten 
via a ``decoupling" as
\begin{eqnarray}
C^\dagger_{x+i}C_{x}+\mbox{H.c}
&=& B_{x+i}^\dagger A_{x+i} A_x^\dagger B_{x}+\mbox{H.c.}\nonumber \\
&\Rightarrow& (B^\dagger_{x+i}W_{xi}B_x+A^\dagger_{x+i}W^\dagger_{xi}
A_{x}+\mbox{H.c.})-|W_{xi}|^2+\cdots.
\label{bfV}
\end{eqnarray}
$W_{xi}$ is an auxiliary complex field  
defined on the link $(x,x+\hat{i})$
where $i=1,2$ is the direction index of the  lattice.
Its  phase degree of freedom $U_{xi}\equiv W_{xi} / |W_{xi}|$
behaves as a {\em spatial component} of 
a compact U(1) gauge field 
$U_{xi} [= \exp(i\theta_{xi})]$. 
$U_{xi}$ represents the  binding force of the constituents $A_x$ and $B_x$.
In the path-integral formalism, the partition function 
${\rm Tr}_C \equiv \exp[-\beta H(C)]\ (\beta\equiv T^{-1})$ 
has the following representation:
\begin{eqnarray}
Z &=& \int \prod_{x,\tau}d\bar{A}_x(\tau)dA_x(\tau) dB_x(\tau)
\prod_{\mu=0,1,2}dU_{x\mu}(\tau)
\exp(A),\nonumber\\
A &=& \int_0^\beta d\tau \sum_{x}
\Big[-i\theta_{x0}-\bar{A}_{x}(\partial_0-i\theta_{x0} +\mu_A)A_x
-\bar{B}_{x}(\partial_0-i\theta_{x0} +\mu_B)B_x\nonumber\\
&&+t\sum_{i}\left(\bar{A}_{x+\hat{i}}U_{xi}(\tau)B_{x} 
+\ {\rm H.c.}\right) + A_{\rm int}\Big ],
\label{nonrel0}
\end{eqnarray}
where
$\tau\ (\in [0,\beta])$ is the continuum imaginary time,
$A_x(\tau)$ is a Grassmann number expressing fermions, 
$\mu_{A(B)}$ is the chemical potential,
and $t$ is the hopping amplitude.
The field $\theta_{x0}$ is a Lagrange multiplier field 
to enforce the constraint (\ref{constraint}).
It may be viewed as the time-component of gauge field.
In fact, in the discrete-time formulation on the 3D lattice 
it appears as the exponent of gauge variable 
in the $\tau$-direction, $U_{x0}=\exp(i\theta_{x0})$.

When the gauge dynamics of $U_{xi}$ is realized in the confinement
phase, the constituents $A_x$ and $B_x$ are bound within electrons
and the relevant low-energy quasi-particles are described by 
the electron operators $C_x$ themselves.
On the other hand, if the gauge dynamics is realized in the deconfinement 
phase, the binding force is weak and these constituents
are no more bound in each electron but dissociate each other.
This is just the the fractionalization(separation) phenomena
of electrons. 

\begin{figure}[htbp]
\begin{center}
\leavevmode
\epsfxsize=9cm     
\epsffile{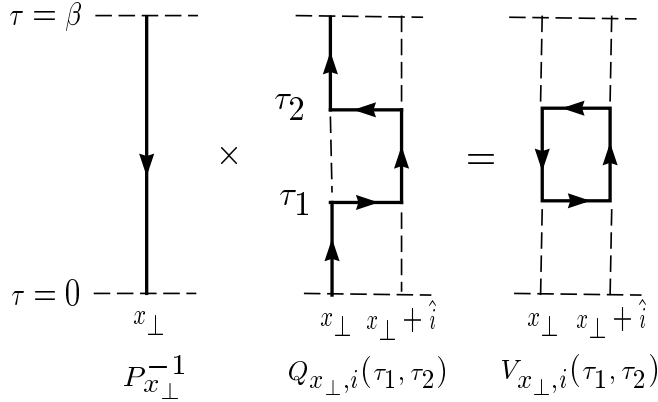}
  \caption{Illustration of the nonlocal interactions
  in electron-fractionalization phenomena. $P_{x_{\bot}}　= \exp(i\int
  d\tau \theta_{x0})$ is the Polyakov
  line at the transverse coordinate
  $x_{\bot} =(x_1,x_2)$, $Q_{x_{\bot},i}(\tau_1,\tau_2)$
 is given by the product of propagators $\langle B_{x_{\bot}}(\tau_1)
 B_{x_{\bot}}^\dagger(\tau_2)\rangle$ 	and  
 $\langle A_{x_{\bot}}(\tau_2)
 A_{x_{\bot}}^\dagger(\tau_1)\rangle$. The product of these $P_{x_{\bot}}$ 
 and $Q_{x_{\bot},i}(\tau_1,\tau_2)$
 results an effective interaction $V_{x_{\bot},i}(\tau_1,\tau_2)$ 
 along a closed loop, which has the same form as
 a vacuum polarization of relativistic matter fields.
}
\label{fig-css}
\end{center}
\end{figure}

In Ref.\cite{pfs} we have studied the system of Eq.(\ref{nonrel0})
at finite temperatures ($T$) by the hopping expansion in the 
temporal gauge $\theta_{x0}=0$.
After integrating over $A_x$ and $B_x$ 
in powers of $t$, one obtains 
the effective interactions.
Up to $O(t^2)$, by restoring the temporal components $U_{x0}$, one gets
the following effective interaction among $U_{x\mu}$,
\begin{eqnarray}
\Delta A &\propto& \delta(1-\delta)
\cdot t^2 \sum_{x,i} \int_0^\beta d\tau 
\int_0^\beta d\tau'\;  
V_{x_{\bot},i}(\tau,\tau') + {\rm c.c.},\nonumber\\
V_{x_{\bot},i}(\tau,\tau')&\equiv& 
\bar{U}_{x_{\bot},\tau',i}\ \exp\big(i\int_\tau^{\tau'} d\tau^{''} 
[\theta_{x_{\bot}+\hat{i},0}(\tau^{''})-
\theta_{x_{\bot},0}(\tau^{''})]\big)\ U_{x_{\bot},\tau,i},
\label{Nonrel}
\end{eqnarray}
where $\delta = \langle A^\dagger_x A_x \rangle$ 
is the concentration of fermions.
The corresponding terms are  illustrated in Fig.\ref{fig-css}.


The interactions among $U_{x\mu}(\tau)$ in $\Delta A$ are quite nonlocal
in the $\tau$-direction, and $\Delta A$ favors the ordered
configurations of $U_{x\mu}$, so the deconfinement phase.
In the previous papers\cite{pfs}, we argued that this is the essence
of the mechanism of fractionalization phenomena like CSS and PFS.
By mapping these gauge models approximately to a spin model, 
we concluded that the U(1) gauge dynamics is 
realized in the deconfinement phase at the low-$T$
region {\em  below} certain  critical line $T_c(\delta)$.
Thus the possible deconfienment phase in the model similar to
Eq.(\ref{Nonrel}) is to support the electron fractionalization phenomena.\\

\subsection{Chern-Simons term by Dirac fermions}

On considering the effects of matter fields upon 3D 
gauge dynamics, there is an 
important difference between fermionic and bosonic matte fields.
That is, relativistic fermions have a possibility to generate 
the Chern-Simons (CS) term $\Delta A_{\rm CS}$ via radiative corrections:
\begin{eqnarray}
\Delta A_{\rm CS} = c \int d^3x\  
\epsilon_{\mu\nu\rho}A_{\mu}\partial_{\nu}A_{\rho},
\end{eqnarray} 
which violates the parity symmetry.
In particular, as the mass term $m_D \bar{\psi}\psi$ of 
3D {\em two-component spinor} Dirac fermion field $\psi$
 violates the parity invariance, it generates the CS term with the coefficient
$c \propto {\rm sgn}(m_D)$ \cite{CST}.
On the other hand, scalar fields do not renormalize the coefficient of
the Chern-Simons term.
In the perturbation theory, the CS term is the leading term 
at long distances and low energies.
So it may change the  phase structure of the gauge 
system, in particular that of the compact gauge systems.
One can intuitively expect appearance of a deconfinement phase since
the CS term suppreses fluctuations of the gauge field.

However in most of the 
effective gauge theories of strongly-correlated electron
systems, the CS term is {\em not spontaneously} generated, 
because the most of these systems including the t-J model preserves 
the parity invariance.
For example, in the flux state of the Heisenberg antiferromagnetic spin model
and the t-J model in the slave fermion representation,
relativistic Dirac fermions appear as low-energy excitations, 
but the parity invariance is preserved in the effective theory 
because the lattice fermions appear in doublets with opposite signatures
of masses\cite{AFH}, which correspond to {\em four-component spinor}
Dirac field in the continuum. 
Then the CS coefficient cancels with each other in the radiative correction
as $c \propto \sum_{m_D = \pm m} {\rm sgn}(m_D) =0$.

Though nonperturbative investigation of the Chern-Simons gauge theory 
is very important by itself\footnote{For example,
phase structure of SU(N) Maxwell-CS
gauge theory has been studied via frustrated Heisenberg spin model 
{\em without} parity invariance, which is a low-energy effective model
of strongly-coupled SU(N) gauge theory of fermions
\cite{IN}. Existence of a deconfinement phase transition is suggested there.}, 
we shall focus on parity-invariant
lattice gauge theory in the rest of discussions in this paper.

\section{Nonlocal U(1) Lattice Gauge Theory}
\setcounter{equation}{0}
In the previous section, we have seen various approaches to study the 
effect of matter fields upon U(1) gauge dynamics.
To confirm these results, one must examine the validity of 
the approximations employed there.
In particular, to check the validity of the hopping expansion at $T=0$ 
is quite important for studies on strongly-correlated electron systems.

Keeping this problem in mind, we start this section with a U(1) LGT 
coupled with matter fields and introduce its effective nonlocal LGT.
Let us consider a 3D cubic lattice (i.e., a 2D lattice 
with a discrete imaginary time). 
The gauge field 
$U_{x\mu}\ (\mu=0,1,2)$ is defined on the link  $(x,x+\hat{\mu})$ 
between the pair of nearest-neighbor sites $x$ and $x+\hat{\mu}$.
The partition function $Z$ is given by the following functional 
integral,
\begin{eqnarray}
Z &=& \int\prod_{x}d\bar{\phi}_x d\phi_x \prod_{x\mu}dU_{x\mu}
\exp(A), \nonumber\\
A &=& -\sum_{x,y}\bar{\phi}_x \Gamma_{xy}(U) \phi_y
+A_U, \nonumber\\
A_U&=& q\sum_{x, \mu < \nu}(\bar{U}_{x\nu}
\bar{U}_{x+\hat{\nu},\mu}U_{x+\hat{\mu},\nu}U_{x\mu}+ c.c.),
\label{original}
\end{eqnarray}
where $x=(x_0,x_1,x_2)$ 
is the site-index of the 3D lattice of the size 
$V = N_0 N_1 N_2$ with the periodic boundary condition, 
$\mu$ is the imaginary-time index ($\mu=0$) and spatial
direction indices ($\mu=1,2$), 
$\phi_x$ is the matter field on $x$,
$U_{x\mu}=\exp(i\theta_{x\mu}) \; (-\pi < \theta_{x\mu} 
\leq \pi)$ is the $U(1)$ gauge variable
on the link $(x,x+\hat{\mu})$,
and $q$ is inverse gauge coupling constant.
$\Gamma_{xy}(U)$ represents the {\it local} minimal
couplings of $\phi_x$ to $U_{x\mu}$.
For example, for a bosonic matter field, 
\begin{equation}
 \sum_{x,y}\bar{\phi}_x \Gamma_{xy}(U) \phi_y
=t\sum_{x,\mu}\left[\bar{\phi}_{x+\hat{\mu}}
U_{x\mu} \phi_x + {\rm H.c.}\right]
+M^2 \sum_x \bar{\phi}_x\phi_x, \; 
 M^2=6+m^2,
\end{equation}
where $m^2$ is the mass in unit of the lattice spacing and $6$ in $M^2$
is the number of links emanating from each site.

After integrating over the matter field $\phi_x$, effective
gauge model is obtained, which includes all contributions from $\phi_x$
to the gauge dynamics,
\begin{eqnarray}
Z &=& \int\prod_{x\mu}dU_{x\mu}
\exp\Big[
f\; {\rm Tr}\;  \log\; \Gamma_{xy}(U) +A_U\Big],
\label{effective-model}
\end{eqnarray}
where $f$ is a parameter counting 
the statistics and internal degrees of freedom of $\phi_x$.
Due to the $({\rm Tr}\;  \log\Gamma_{xy}(U))$ term, the effective gauge 
theory becomes {\it nolocal}. 
For relativistic matter fields, a formal expression of the
effective gauge theory action is obtained by the hopping expansion and
it is expanded as a sum over
all the closed random walks ${\cal R}$ (loops including
backtrackings) on the 3D lattice, which  represent world lines 
of particles and antiparticles as
\begin{equation}
{\rm Tr}\;  \log\; \Gamma_{xy}(U)= \sum_{\cal R}
\frac{\gamma^{L[{\cal R}]}}{L[{\cal R}]}
\prod_{(x\mu) \in {\cal R}}U_{x\mu}.
\label{eff1}
\end{equation}
$L[{\cal R}]$ is the length 
of ${\cal R}$, and $\gamma =(6+m^2)^{-1}$ is the hopping parameter.
There are many different random walks that have the same shape of  a 
closed loop on the lattice.
Each random walk in such a family may have a different starting point
and/or backtrackings. This degeneracy
cancels out the denominator $L[{\cal R}]$ in Eq.(\ref{eff1}). 
For the constant gauge-field configuration $U_{x\mu}=1$, the expansion
in (\ref{eff1}) is logarithmically divergent $\sim\log m$
as $m \rightarrow 0$
due to the lowest-energy zero-momentum mode. 

\begin{figure}[htbp]
\begin{center}
\leavevmode
\epsfxsize=5cm     
\epsffile{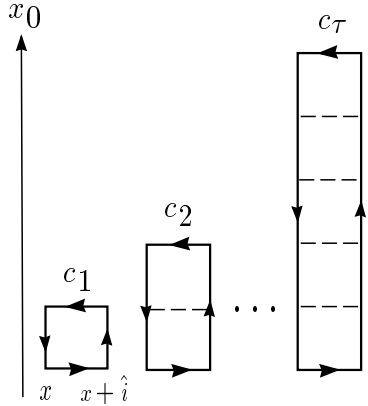}
  \caption{Illustration of the nonlocal interaction in the action
  (\ref{simplified-model}). Each rectangle with thick lines represents
  $V_{x,i,\tau}.$
}
\label{fig-nonlocalint}
\end{center}
\end{figure}

Below we shall study a slightly more tractable model than  
that given by Eq.(\ref{effective-model}). It is suggested by
the hopping expansion (\ref{eff1}),
and  obtained by retaining only the rectangular loops  
extending in the $\tau$-direction in the loop sum 
and choosing their expansion coefficients as follows;
\begin{eqnarray}
Z_{\cal T} &=& \int\prod_{x\mu}dU_{x\mu}
\exp(A_{\cal T}), \nonumber\\
A_{\cal T} &=& g\sum_{x}\sum_{i=1}^2
\sum_{\tau=1}^{N_0} c_{\tau}(V_{x,i,\tau}
+\bar{V}_{x,i,\tau})
+A_S,\nonumber\\
V_{x,i,\tau} &=& \bar{U}_{x+\tau\hat{0},i}
\prod_{k=0}^{\tau-1}\left[\bar{U}_{x+k\hat{0},0}
U_{x+\hat{i}+k\hat{0},0}\right]U_{xi},\nonumber\\
A_S &=& g \lambda\sum_x(\bar{U}_{x2}
\bar{U}_{x+\hat{2},1}U_{x+\hat{1},2}U_{x1}+ {\rm c.c.}),
\label{simplified-model}
\end{eqnarray}
where $g$ is the (inverse) gauge coupling constant, and 
$V_{x,i,\tau} $ 
is the product of $U_{x\mu}$ along the rectangular
$(x, x+\hat{i}, x+\hat{i}+\tau \hat{0},x+ \tau \hat{0})$ 
of size $(1 \times \tau)$ in the $(i-0)$ plane. 
See Fig.\ref{fig-nonlocalint}.


In $A_S$, we have retained only the single-plaquette coupling
with the coefficient $(g\lambda)$.
For the nonlocal coupling constant $c_{\tau}$,
we consider the following three cases; 
\begin{eqnarray}
c_{\tau} &=& \left\{
\begin{array}{ll}
\tau^{-\alpha}, \ & 
{\rm power-law\ decay\ (PD-}\alpha)\ (\alpha=1,2,3), \\
e^{-m \tau}, & {\rm exponential\ decay\  (ED)}, \\
1, &{\rm no\ decay\  (ND)}.   \\
\end{array}
\right.
\label{ctau}
\end{eqnarray}
The power $\alpha=1$ in the PD model in (\ref{ctau})
reflects the effect of the {\em relativistic} massless excitations
without dimensional parameters. 
In fact, this $c_{\tau}$ generates 
a logarithmically divergent action for $U_{x\mu}=1$
explained below Eq.(\ref{eff1}) 
 as one can see from the relation,
$\sum_\tau \exp(-m \tau)\tau^{-1} \simeq \log (1/m)$.
The action for $m=0$ is then proportional to
$\sum_\tau \tau^{-1} \simeq \log N_0$ for finite $N_0$. 
On the other hand, the ED model contains the parameter
$m$ with mass dimension and simulates the 
case of massive matter fields.
(We used Eq.(\ref{ctau}) instead of
$\exp(-m\tau)/\tau$ to make the comparison with the PD case more 
definitive.)
The ND model corresponds to gauge model coupled to {\em nonrelativistic}
fermions with a Fermi surface (or Fermi line) as it is seen from
Eq.(\ref{Nonrel}).

\section{Numerical Results}
\setcounter{equation}{0}

In this section, we report our numerical calculations of 
the internal energy, the specific heat, the Polyakov lines, etc., 
and determine the phase structure of the models.
Most of our interest concerns a possible 
CDPT in the 3D compact U(1) gauge
models with the nonlocal interactions.

In the MC simulations, we consider the isotropic lattice,
$N_\mu = N\;(\mu=0,1,2)$,  with the periodic boundary condition up to
$N = 32$, where the limit $N\rightarrow  \infty$ corresponds to the
system on a 2D spatial lattice at $T=0$.
For the mass of the ED model,
we set $m=1$. 
For the spatial coupling $\lambda$
scaled by $g$, we consider the two typical cases
$\lambda=0$ (i.e., no spatial coupling) and $\lambda=1$.

\subsection{Internal energy and specific heat}

First we calculate the
following ``internal energy" 
$E$ and the ``specific heat" $C$ per site;
\begin{eqnarray}
Z_{\cal T}&=&\int[dU]\exp(A_{\cal T})\equiv \exp(-FV),\nonumber\\
E &\equiv& -\frac{1}{V}\langle A_{\cal T}\rangle = -\frac{1}{V}
\frac{g}{Z_{\cal T}}
\frac{dZ_{\cal T}}{dg}= g\frac{dF}{dg}, \nonumber\\
C &\equiv& \frac{1}{V}\langle (A_{\cal T}
- \langle A_{\cal T}\rangle)^2\rangle = -g^2\frac{d^2F}{dg^2}.
\label{EC}
\end{eqnarray}  
Here we note that the definition of $C$ (\ref{EC}) is 
the response of $E$ under the variation of $g$ and is different from
the conventional specific heat that measures the response
under the variation of temperature itself.
The latter contains extra terms associated with the 
change of $N_0$. Because  these terms behave less singular 
than the variance of $E$, they are irrelevant in searching for
 phase transitions.


\begin{figure}[htbp]
\begin{center}
\leavevmode
\epsfxsize=9cm     
\epsffile{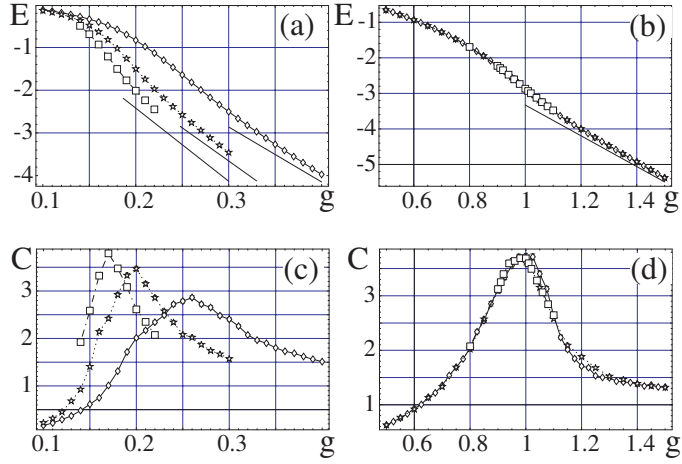}
  \caption{Internal energy $E$ and its fluctuation $C$ of the action  
with $\lambda=1$ vs non-local coupling $g$ 
for $N =8 (\blacklozenge), 16(\bigstar), 24(\blacksquare)$;
(a,c) PD-1 model, (b,d) ED model.
The solid lines in (a) and (b) are the leading order in the 
large-$g$ expansion (LTE).
In the PD-1 model, strong $N$ dependence is observed in $E$ 
at large $g$ and in the developing peak of $C$. 
They indicate a second-order phase transition in the PD-1 model.
In the ED model, on the other hand, $C$ does not develop and therefore
the observed peak indicates not a phase transition but a crossover.
}
\label{fig-EC}
\end{center}
\end{figure}

In Fig.\ref{fig-EC}, we present $E$ and $C$ for $\lambda=1$ vs the
nonlocal coupling $g$.
In Appendix we calculate $E$ and $C$ by the  high-temperature expansion (HTE) 
for small $g$ and   by the  low-temperature expansion (LTE) 
for large $g$. 
The MC results of $E$ and $C$
are consistent with these expansions.
For comparison with HTE (e.g., for  $0 < g < 0.1$ in the PD-1 model ($\alpha=1$)), see  Fig.\ref{fig-hte-mc} in Appendix. 
For comparison with LTE, the leading result of LTE 
is  shown by the straight lines in Fig.\ref{fig-EC}(a,b)
and $C=1$ in  Fig.\ref{fig-EC}(c,d).

First, let us see the PD-1 model ($\alpha=1$) in detail.
 $E$ of Fig.\ref{fig-EC}(a) connects 
the  results of HTE and LTE. 
Since the LTE is an expansion around $V_{x,i,\tau} =1$,
this behavior implies $V_{x,i,\tau} \sim 1$ for large $g$.
$C$ of Fig.\ref{fig-EC}(c) shows that its peak develops 
as the system size $N$ increases.  
These two points indicate that the PD-1 model exhibits
{\em a second-order phase transition separating the disordered
(confinement) phase and the ordered (deconfinement) phase}
at $g=g_c \simeq 0.17$ which is determined from the data of  $N=24$. 
The existence and the nature of 
the phase transition will be confirmed later by the measurement of
the Polyakov lines and the instanton density as we show in the 
following subsections.

We notice that the location of the peak in $C$ of Fig.\ref{fig-EC}(c)
shifts to {\it smaller} $g$ as the system size $N$ increases
in the direction opposite to the usual second-order phase transitions.
This behavior reflects the fact that the couplings among
$U_{x\mu}$ increases effectively as $N$ increases
due to the additional terms in the summation over $N_0$ in the action
even if one fixes the overall constant $g$.
This is the characteristic nature of nonlocal interactions
in strong contrast to local interactions.

On the contrary, in the ED model of Fig.\ref{fig-EC}(d), 
the peak of $C$ does not develop as $N$ increases,
showing  {\em no} signals of a second-order transition. 
It may have a higher-order transition or just  a {\em crossover}.
Similar behavior of $C$ is observed in the ordinary U(1) gauge systems
with local actions which have only the confinement phase.
The physical meaning of the above ``crossover" in the ED model 
shall become clear by studying instantons in Sect.4.3. 

\begin{figure}[htbp]
\begin{center}
\leavevmode
\epsfxsize=7cm     
\epsffile{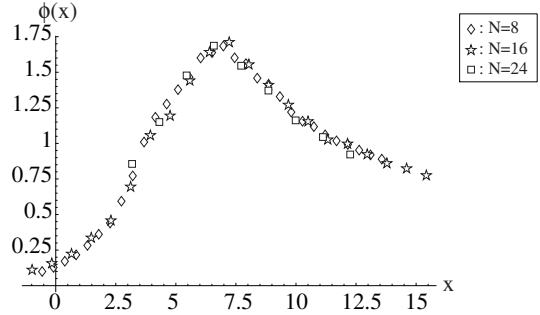}
  \caption{Scaling function $\phi(x)$ of (\ref{FSS})
  obtained from the data $C$ of the PD-1
model in Fig.\ref{fig-EC}(c) for $N=8,16$ and $24$.  
The result obviously shows that the finite-size scaling law holds and 
the observed phase transition in the PD-1 model is of second-order.
}
\label{fig-FSS}
\end{center}
\end{figure}

It is quite interesting to see whether the data of $C$ for 
$N=8, 16$ and $24$ in Fig.\ref{fig-EC}(c) exhibit the 
finite-size scaling behavior\cite{FSS}.
To this end,
let us introduce a parameter 
\begin{eqnarray}
\epsilon\equiv (g-g_\infty)/g_\infty
\end{eqnarray}
 where
$g_\infty$ is the critical gauge coupling of the infinite system at
$N\rightarrow \infty$.
Then let us assume that
the correlation length $\xi$ scales as $\xi \propto \epsilon^{-\nu}$
with a critical exponent $\nu$.
We also expect that
the peak of $C$ diverges as $C_{\rm peak}\propto \epsilon^{-\sigma}$
as $N \rightarrow \infty$ with another critical exponent $\sigma$.
The finite-size scaling hypothesis predicts that the specific heat 
$C(\epsilon,N)$ for sufficiently large $N$ scales as 
\begin{equation}
C(\epsilon,N)=N^{\sigma/\nu}
\phi(N^{1/\nu}\epsilon),
\label{FSS}
\end{equation}
where $\phi(x)$ is a certain scaling function.
In Fig.\ref{fig-FSS}, we present $\phi(x)$ determined by
using the data in Fig.\ref{fig-EC}(c)
with $\nu=1.2, \; \sigma/\nu=0.25$ and $g_\infty=0.11.$
This result 
indicates that the finite-size scaling law holds quite well.
Cbonsidering the errors in the data we estimate the values of
scaling parameters and $g_\infty$ as
\begin{eqnarray} 
\nu=1.2\sim 1.3, \; \sigma/\nu=0.25\sim 0.26,\ g_\infty=
0.10\sim 0.12.
\end{eqnarray}


The simulations of the PD-1 and ED models with $\lambda=0$ 
give similar behaviors of $E$ and $C$ as the $\lambda=1$ case, 
preserving the above phase structure for $\lambda=1$.
That is, the PD-1 model ($\lambda=0$) has a CDPT, while
the ED model ($\lambda=0$) has no transition.

\begin{figure}[htbp]
\begin{center}
\leavevmode
\epsfxsize=10cm     
\epsffile{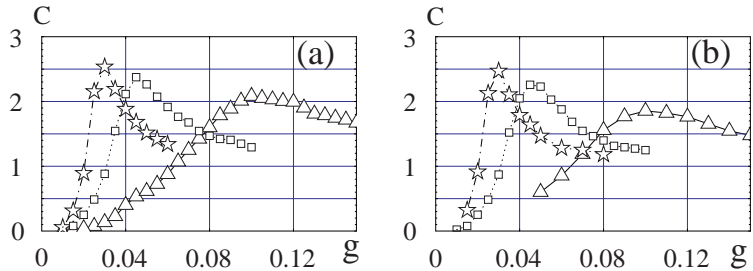}
  \caption{ $C$ vs $g$ in the ND model with (a) $\lambda=0$ 
and (b) $\lambda=1$ for  $N =8 (\blacktriangle), 16(\blacksquare), 
24(\bigstar)$. The peaks develops as the system-size increases.
}
\label{fig-ND}
\end{center}
\end{figure}

Let us next consider the ND model.
In  Fig.\ref{fig-ND}, we present $C$ for $\lambda=0$ and 1,
which show strong signals of a  second-order
phase transition.
However, the value of the critical coupling for $\lambda=1$,
$g_c \sim 0.1 (N=8), 0.045 (N=16), 0.03 (N=24)$, 
decreases very rapidly as the system size increases.
This behavior is explained by the increase of effective coupling
explained above.
Then one may think that $g_c \rightarrow 0$ as $N\rightarrow \infty$,
i.e., only the deconfinement phase survives in the ND model.
This expectation is consistent with the fact that 
the coefficient $Q_2\equiv\sum_\tau c^2_\tau$ of  
HTE, $C=2(2Q_2+\lambda^2)g^2+O(g^3)$ in  
Eq.(\ref{hteec}) diverges as $Q_2 \propto N \rightarrow
\infty $ for the ND model,
i.e., the  radius of convergence in the HTE is zero.
On the contrary, $Q_2$ is finite for the PD-1 model, assuring
us $g_c \neq 0$. This is supported also by the scaling analysis
given above.


\begin{figure}[htbp]
\begin{center}
\leavevmode
\epsfxsize=9cm     
\epsffile{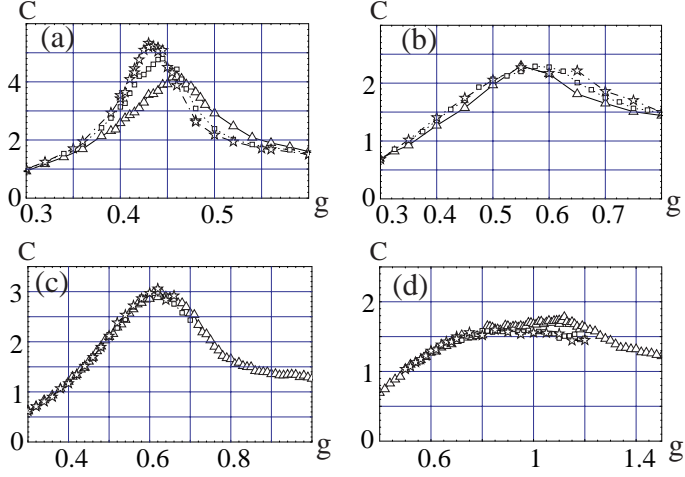}
  \caption{$C$ vs $g$ in the PD-2 and PD-3 models
for  $N =8 (\blacktriangle), 16(\blacksquare), 24(\bigstar)$. 
(a) PD-2($\lambda=1$),
(b) PD-2($\lambda=0$), (c) PD-3($\lambda=1$), (d) 
PD-3($\lambda=0$). 
The peak in $C$ develops in (a) as $N$ 
increases, whereas it does not in (b), (c) and (d).  
This indicates that 
the CDPT starts to appear between the two cases, 
(a) PD-2($\lambda=1$) and (b) PD-2($\lambda=0$).
}
\label{fig-PD-23}
\end{center}
\end{figure}

From these results for the various cases of long-range interaction,
it seems that there exists a {\em critical power} 
$\alpha = \alpha_c$ below which the CDPT takes place.
In Fig.\ref{fig-PD-23}
we show $C$ of the PD model with $\alpha=2$ and
$\alpha=3$.


We obtain a very interesting result, i.e., 
for the PD-2 system ($\alpha=2$) with the {\em nonvanishing} 
spatial coupling
$\lambda=1$, there exists a second-order CDPT as in the PD-1 case,
whereas in the PD-3 case ($\alpha=3$) the peak in $C$ does not develop
as $N$ increases, hence no signals of CDPT.
Furthermore, careful study of the PD-2 case shows that the second-order
CDPT {\em disappears} for {\em vanishing} spatial coupling 
$\lambda=0$. 
Thus we conclude that the critical power is $\alpha_c=2$
and the spatial coupling $\lambda$ controls the existence
of the CDPT.
The above results will be confirmed by the study of the Polyakov line
in the following subsection.


\subsection{Polyakov lines}

We  have argued the possible CDPT by 
measuring the thermodynamic quantities like $E$ and $C$.
In order to study the CDPT in more details and further
the nature of gauge dynamics in each phase, it is useful 
to calculate order parameters in gauge theory like 
Polyakov lines and Wilson loops.

First, let us introduce the Polyakov lines $P_{x_{\bot}}$
for each spatial site $x_{\bot}\equiv (x_1,x_2)$ and study their 
spatial correlations  $f_P(x_{\bot})$,
\begin{eqnarray}
&& P_{x_{\bot}} = \prod_{x_0=1}^{N_0}U_{x_{\bot},x_0,0},\ 
x_{\bot}=(x_1,x_2), \nonumber \\
&& f_P(x_{\bot}) = \langle \bar{P}_{x_{\bot}} P_{0}\rangle.
\label{Plines}
\end{eqnarray}
Since the present model (\ref{simplified-model}) contains
no long-range interactions in the spatial directions, 
$f_P(x_{\bot})$ is expected to supply a good order parameter
to detect a possible CDPT.
In the deconfinement phase, the fluctuations 
of $U_{x0}$ are small, which implies an  order in $f_P(x_{\bot})$,
i.e., we expect $f_P(x_{\bot})\neq 0$ as $x_{\bot} \rightarrow$
large in the deconfinement phase.

\begin{figure}[htbp]
\begin{center}
\leavevmode
\epsfxsize=10cm     
\epsffile{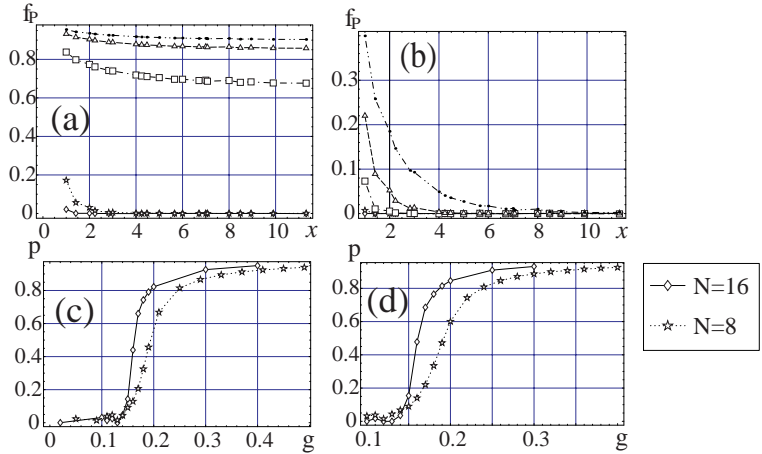}
\caption{Correlations of Polyakov lines, $f_P(x_{\bot})$,
vs $|x_{\bot}|$.
(a) PD-1 ($\lambda=0$, N=16) with $g= 0.4,0.3,0.2,0.1,0.02$ from above.
(b) ED ($\lambda=0$, N=16) with $g= 2.5,2.0,1.5,1.0,0.5$ from above.
In (c) and (d) the order parameter $p=(f_P(x_{\bot}^{\rm MAX}))^{1/2}$ 
vs $g$ is plotted for the PD-1 model; (c) PD-1 $\lambda=0$ and  
(d) PD-1 $\lambda=1$. 
They  exhibit  long-range orders for $g > g_c \simeq 0.15$
  in the PD-1 model for both $\lambda$.
}
\label{fig-P}
\end{center}
\end{figure}

In Fig.\ref{fig-P}, we  present $f_P(x_{\bot})$ for the PD-1 and 
ED models.
The PD-1 model of Fig.\ref{fig-P}(a) clearly exhibits an off-diagonal 
long-range order,
i.e., $\lim_{x_\bot \rightarrow \infty} f_P(x_{\bot}) \neq 0 $
for $g \geq 0.20$, whereas the ED model of Fig.\ref{fig-P}(b) {\it does not} 
for all 
$g$'s.
To see this explicitly, we plot in Figs.\ref{fig-P}(c) and (d) 
the order parameter $p 
\equiv (f_P(x_{\bot}^{\rm MAX}))^{1/2} $ for the PD-1 model, 
where $x_{\bot}^{\rm MAX} \equiv N/\sqrt{2}$ is the 
distance at which $f_P$ becomes minimum 
due to the periodic boundary condition.
$p$ of the PD-1 model($\lambda=0$) 
starts to develop continuously from zero  
at $ g = g_c \simeq 0.15$. The size dependence of $p$
shows a typical behavior of a second-order phase transition.
Thus  the gauge dynamics of the PD-1 model is 
realized in the deconfinement phase for $g > g_c$,
whereas it is in the confinement phase for $g < g_c$. 
In contrast, the ED model stays 
always in the confinement phase.
These results including the value of $g_c$ are in good agreement with 
those derived from the data of $E$ and $C$ given in Fig.\ref{fig-EC}.

\begin{figure}[htbp]
\begin{center}
\leavevmode
\epsfxsize=12cm     
\epsffile{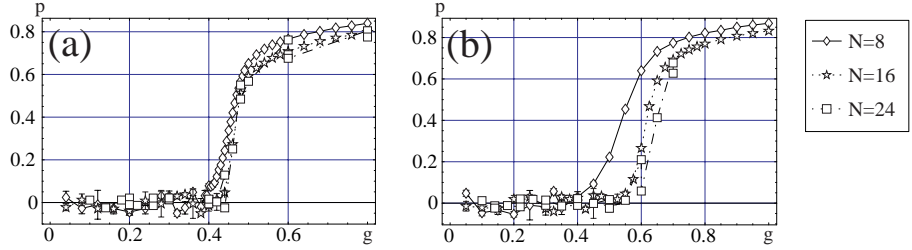}
\caption{
Off-diagonal long-range order of Polyakov lines, $p$ 
vs $|x_{\bot}|$ for the PD-2 model with (a) $\lambda=1$ and 
(b) $\lambda=0$. 
 In the $\lambda=1$ case, $p$
exhibits a steeper jump for larger $N$, whereas
it does not in the $\lambda=0$ case.
}
\label{fig-PD2P}
\end{center}
\end{figure}

Let us turn to the PD-2 model with and without the spatial coupling.
In Fig.\ref{fig-PD2P}, we show the result of $p$ for $\lambda=1$ and 
$\lambda=0$.
We observe that the model with $\lambda=1$ shows a typical
behavior of the second-order phase transition as the system size is 
increased, whereas the case of $\lambda=0$ does not.
From this result and the observation of $C$ in the previous
subsection, we conclude that the CDPT exists in the PD-2 model 
$(\lambda=1)$ whereas it disappears in 
the PD-2 model $(\lambda=0)$.

\begin{figure}[htbp]
\begin{center}
\leavevmode
\epsfxsize=10cm     
\epsffile{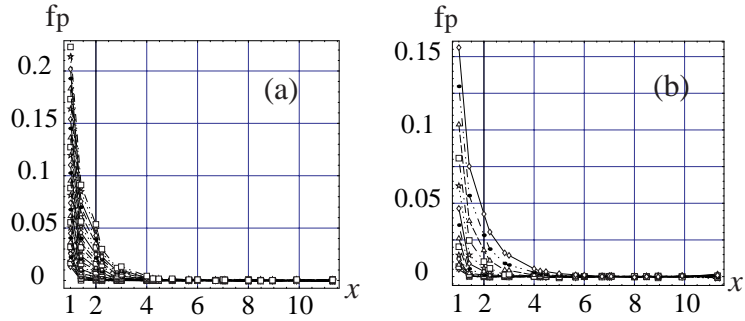}
\caption{
$f_P(x_{\bot})$ vs $g$ for the PD-3 model with $N=16$;
(a) $\lambda=1$ for $g=0.4 \sim 0.7$, (b) $\lambda=0$ for
$g=0.56 \sim 0.78$. The upper curves have
larger $g$'s.
It is obvious that
there is no long-range order in the PD-3 model regardless
of the values of $g$ and $\lambda$.
}
\label{fig-pl-d3}
\end{center}
\end{figure}

In Fig.\ref{fig-pl-d3} we present  $f_P(x_{\bot})$ of the PD-3 model. 
It is obvious that there is no long-range order 
in the PD-3 model both for $\lambda=0,1$ and 
only the confinement phase is realized
for all $g$. 

\subsection{Wilson loops}

Let us turn to study of the Wilson loops.
For ordinary {\it pure and local} gauge systems, the Wilson loop 
$W[{\cal C}]$ along
a closed loop ${\cal C}$ on the lattice is 
a good order parameter to study the gauge dynamics;
$W[C]$ obeys the area law in the confinement phase and
the perimeter law in the deconfinement phase; 
\begin{eqnarray}
W[{\cal C}] \equiv \langle \prod_{\cal C} U_{x\mu} \rangle \sim
\left\{
\begin{array}{ll}
\exp(-a S[{\cal C}]),\ & {\rm area\ law}, \\ 
\exp(-a' L[{\cal C}]),\ & {\rm perimeter\ law},
\end{array}
\right.
\label{wilson-loop}
\end{eqnarray}
where $S[{\cal C}]$ is the minimum area of a surface, the boundary of 
which is ${\cal C}$, and $a$ and $a'$ are constants.
For a (local) gauge theory containing matter fields of the 
fundamental charge, 
$W[{\cal C}]$ cannot be an order parameter because the 
matter fields generate the terms
$\prod_{\cal C} U_{x\mu}$ 
with coefficients $\sim \exp(-b L[{\cal C}])$ in the effective action. 
However, in the present model (\ref{simplified-model}), 
the nonlocal terms are restricted only along
the temporal direction, so it is interesting to measure
$W[\cal C]$ for the loops 
lying in the {\it spatial (1-2) plane}.
If $W[\cal C]$ obeys the perimeter law, fluctuations of the spatial
component of  gauge field is small.

\begin{figure}[htbp]
\begin{center}
\leavevmode
\epsfxsize=12cm     
\epsffile{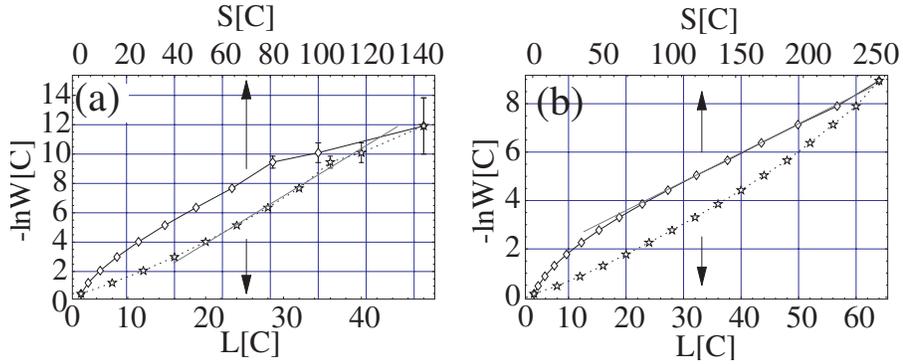}
\caption{Wilson loops ($N=32$) in the 1-2 plane
at large $g$  vs $L[\cal C]$ or $S[\cal C]$. 
(a) PD($\lambda=1,g =0.25$), (b) ED($\lambda=1,g= 1.5$). 
The PD model seems to prefer the perimeter law,
whereas the ED model prefers the area law.}
\label{fig-W}
\end{center}
\end{figure}

In Fig.\ref{fig-W}, we plot $W[{\cal C}]$. 
For the PD-1 model in Fig.\ref{fig-W}(a), the data at $g=0.25$ 
seem to prefer
 the perimeter law. For the ED model in Fig.\ref{fig-W}(b), 
the area law fits $W[{\cal C}]$  
better than the perimeter law at $g = 1.5$;
a considerably larger value than $g \simeq 1.0$ at the peak
of $C$. This suggests that the area law holds in the ED model
at all $g$. These observations are consistent with the previous results
on the (non)existence of the CDPT.
We conclude that the Wilson loops in the spatial plane
can be used as  an order parameter of gauge dynamics in the present model.

The result that the PD-1 case 
of the simplified model (\ref{simplified-model}) exhibits the
CDPT strongly suggests 
that the original model (\ref{effective-model}) of massless matter fields 
also has the deconfinement phase, 
because the isotropic distribution of the nonlocal gauge couplings 
in the original model should give the similar effect
of suppression of fluctuations of $U_{x\mu}$ as those in the 
temporal direction in Eq.(\ref{simplified-model}).
This expectation is supported by the measurement of the spatial Wilson
loop given above.

\subsection{Instantons}

It is well known in a 3D continuum space-time that 
instanton (monopole) configurations of U(1) gauge field carry
nontrivial topological numbers, and the instanton density
serves as an index to
express the disorderness of gauge field\cite{polyakov}.
To see the details of th gauge dynamics of the present model
and to support the conclusions 
obtained in the previous subsections, let us study instantons on the lattice.
We employ the definition of the instanton density $\rho_x$ 
at the site $x$  in U(1) lattice gauge theories by
DeGrand and Toussaint \cite{instanton}. 
We introduce the ``vector potential" $\theta_{x\mu}$ as the exponent
of $U_{x\mu}=\exp (i\theta_{x\mu})$ $[\theta_{x\mu}\in (-\pi,\pi)]$.
Then, the magnetic flux $\Theta_{x,\mu\nu}$
penetrating plaquette $(x,x+\mu,x+\mu+\nu,x+\nu$) is expressed as 
\begin{eqnarray}
&& \Theta_{x,\mu\nu}\equiv \theta_{x\mu}+\theta_{x+\mu,\nu}
-\theta_{x+\nu,\mu}-\theta_{x\nu}, \ \ (-4\pi<\Theta_{x,\mu\nu}<4\pi).
\label{Theta}
\end{eqnarray}
We decompose $\Theta_{x,\mu\nu}$ into its 
{\it integer} part $2\pi n_{x,\mu\nu}$ ($n_{x,\mu\nu}$ is an integer)
and the remaining part  $\tilde{\Theta}_{x,\mu\nu} \equiv$
 $\Theta_{x,\mu\nu}\; (\mbox{mod} \;2\pi$) uniquely,
\begin{equation}
\Theta_{x,\mu\nu}=2\pi n_{x,\mu\nu}+\tilde{\Theta}_{x,\mu\nu}, \;\;
(-\pi<\tilde{\Theta}_{x,\mu\nu}<\pi).
\end{equation}
Physically speaking, $n_{x,\mu\nu}$ describes the 
Dirac string whereas $\tilde{\Theta}_{x,\mu\nu}$ describes the
fluctuations around it.
The quantized instanton charge $\rho_x$ at the cube 
around the site $\tilde{x} =x+(\hat{1} +\hat{2} +\hat{3})/2$ of the dual lattice 
is defined as 
\begin{eqnarray}
\rho_x&=&
-{1\over 2}\sum_{\mu,\nu,\rho}\epsilon_{\mu\nu\rho}
(n_{x+\mu,\nu\rho}-n_{x,\nu\rho})\nonumber\\
&=&{1\over 4\pi}\sum_{\mu,\nu,\rho}\epsilon_{\mu\nu\rho}
(\tilde{\Theta}_{x+\mu,\nu\rho}-\tilde{\Theta}_{x,\nu\rho}),
\label{instden}
\end{eqnarray}
where $\epsilon_{\mu\nu\rho}$ is the complete antisymmetric tensor.
$\rho_x$ measures the total flux emanating
from the monopole(instanton) sitting at $\tilde{x}$. 
Roughly speaking, $\rho_x$ measures the strength of nonperturbative
gauge fluctuations around $\tilde{x}$.
For the {\it local} 3D U(1) compact lattice gauge 
theory without matter fields,
the average density 
\begin{eqnarray}
\rho \equiv \frac{1}{V}\sum_x\langle |\rho_x| \rangle,
\label{rho}
\end{eqnarray} 
is known to 
behave as $\rho \propto \exp(-cg)$ ($c$ is a constant)
if the instanton action $cg$ is large; The instanton gas
stabilizes the confinement phase for all the gauge coupling\cite{polyakov}.

\begin{figure}[htbp]
\begin{center}
\leavevmode
\epsfxsize=9cm     
\epsffile{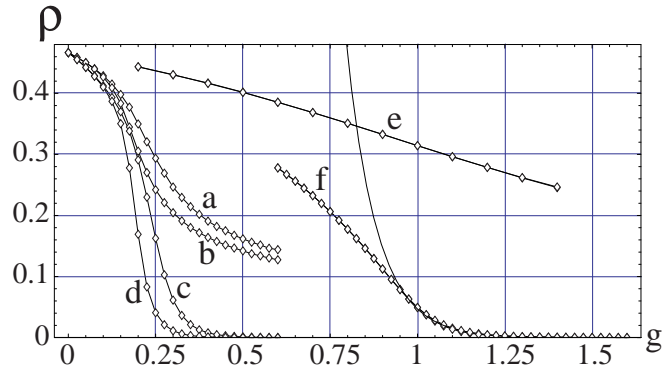}
\caption{
Average instanton density $\rho$ vs $g$.
(a)PD-1($\lambda=0$, $N=8$), (b)PD-1($\lambda=0$, $N=16$),
(c)PD-1($\lambda=1$, $N=8$), (d)PD-1($\lambda=1$, $N=16$),
(e)ED($\lambda=0$, $N=8,16$),
(f)ED($\lambda=1$, $N=8,16$).
The solid curve is $\propto \exp(-cg)$ and fits (f) at large $g$. 
}
\label{fig-Ins}
\end{center}
\end{figure}

In Fig.\ref{fig-Ins} we present
$\rho$ as a function of $g$ in the PD-1 and ED models. 
It decreases as $g$ increases  more rapidly in 
the PD-1  model than in the ED model. 
This difference in behavior is consistent with the result 
that the PD-1 model exhibits a second-order transition, while
the ED model does not.
The $\lambda$ coupling
enhances the rate of decrease in $\rho$ as one expects
since the spatial coupling enhances the ordered
deconfinement phase. 
In the ED model 
with $\lambda=1$, $\rho$ is fitted by $\exp(-cg)$ 
in the dilute (large $g$) 
region, and the smooth increase for smaller $g$
indicates  a {\em crossover} from 
the {\em dilute} gas of instantons 
to the {\em dense} gas, just the  behavior similar to the case of
pure and local lattice gauge theory\cite{polyakov,crossover}. 

In Ref.\cite{AIMS} we presented the figure (Fig.3) which is
very similar to Fig.\ref{fig-Ins}. However, the former was plotted by using
the definition $\rho \equiv V^{-1}\sum_x \langle (1-\delta_{0,\rho_x})  
\rangle$, the average of occupation number of instantons
(1 for $\rho_x \neq 0$ and 0 for $\rho_x =0$) per site 
instead of the instanton density of Eq.(\ref{rho}) itself. 
The curves in two figures are almost 
indistinguishable, because the configirations with $|\rho_x| \geq 2$ are 
rather rare.

\begin{figure}[htbp]
\begin{center}
\leavevmode
\epsfxsize=9cm     
\epsffile{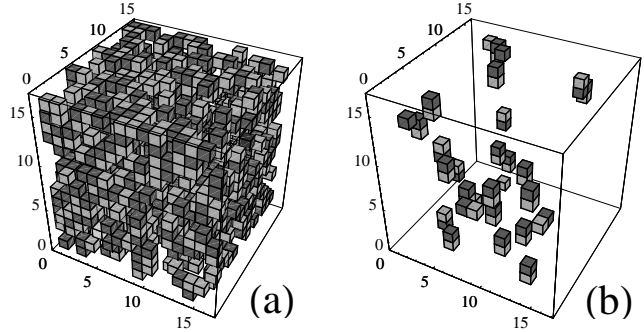}
\caption{
Snapshots of instanton configuration $\rho_x$ on the $16^3$
lattice for 
(a) PD-1($\lambda=1, g= 0.15$), and (b) PD-1($\lambda=1, g= 0.30$).
The light cubes denote $\rho_x=1$ and the dark cubes 
$\rho_x=-1$. }
\label{fig-SS}
\end{center}
\end{figure}

In Fig.\ref{fig-SS} we present snapshots of $\rho_x$ 
for the PD-1 model with $\lambda=1$. Fig.\ref{fig-SS}(a) is a dense gas
whereas Fig.\ref{fig-SS}(b) is a dilute gas. They are separated at 
$g_c \simeq 0.20$, the location of the 
peak of $C$ for $N=16$.
In Fig.\ref{fig-SS}(b), instantons mostly appear in dipole pairs 
at nearest-neighbor sites, $\rho_x=\pm 1, \rho_{x\pm \mu}=\mp 1$,
 while in Fig.\ref{fig-SS}(a), they appear densely and it is 
 hard to determine their partners.
In both cases, the distributions $\rho_x$ have no apparent
 anisotropies like column structures. However, the orientations
of dipoles in Fig.\ref{fig-SS}(b) are mostly ($\sim 92\%$) in the 
temporal direction 
as expected from the nonlocal interactions of 
Eq.(\ref{simplified-model}).

\begin{figure}[htbp]
\begin{center}
\leavevmode
\epsfxsize=12cm     
\epsffile{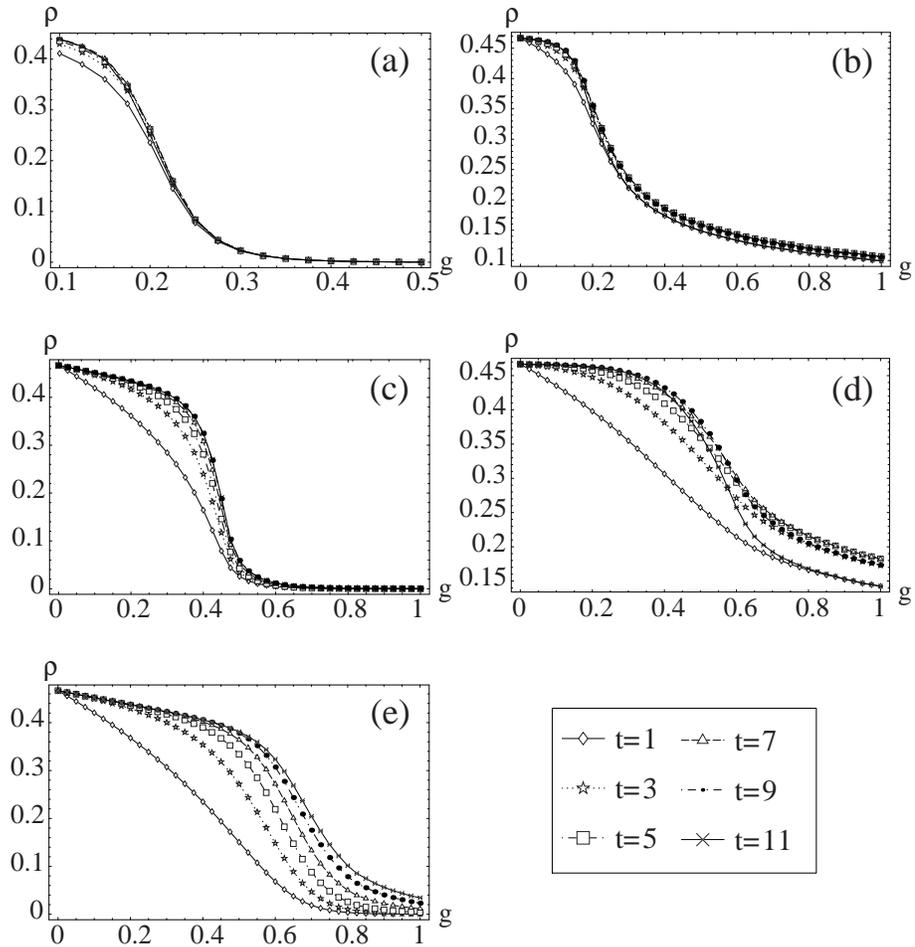}
\caption{
Density of nonlocal instantons $\rho_t$  vs $g$ for $N=12$.
(a) PD-1($\lambda=1$), (b) PD-1($\lambda=0$),
(c) PD-2($\lambda=1$), (d) PD-2($\lambda=0$),
(e) PD-3($\lambda=1$). $\rho_t$ at a fixed $g$ increases
as $t$ increases.
 }
\label{fig-rho-size}
\end{center}
\end{figure}

Next, let us examine  the difference of gauge dynamics
in various cases of nonlocal interactions in detail.
To this end, it is useful to measure new observables made out
of gauge-field configurations;
nonlocal instantons elongated  in the temporal direction
with the length $t =2,3,\cdots,N-1$. 
The density of these instantons is defined by
\begin{eqnarray}
\rho_{x,t}&\equiv& \sum_{\tau=0}^{t-1}\rho_{x+\tau \hat{0}},
\nonumber\\
\rho_t &\equiv& \langle \rho_{x,t} \rangle.
\label{nonlocalinstden}
\end{eqnarray}
In Fig.\ref{fig-rho-size} we plot $\rho_t$ vs $g$ for various cases
of the exponent $\alpha$ and $\lambda$.

We notice that $\rho_t$'s in Fig.\ref{fig-rho-size}(a,b,c)
behave quite differently from $\rho_t$'s in Fig.\ref{fig-rho-size}(d,e).
For the former cases, all the curves of $\rho_t$ with different $t$'s  
decrease similarly as $g$ increases [except for $t=1$ in (c)],
whereas for the latter cases, each $\rho_t$ decreases in a  different manner. 
The present nonlocal instanton density $\rho_t$  is capable to
distinguish the cases (a,b,c) exhibiting a CDPT and 
the cases (d,e) without transition (cross over) through its $t$ dependence.
This is consistent with our common understanding that 
the second-order phase transition is
a collective phenomenon at which wild fluctuations of all the variables
in the system in the disordered phase start to be reduced coherently 
and also scale-invariantly near the transition point.


\section{Effective Models at Large $g$ and $\lambda = 0$}
\setcounter{equation}{0}

In this section, we study a 1D 
XY spin model for the {\em spatial} gauge variables $U_{xi}$ and also
a 2D XY spin model for the {\em temporal} ones $U_{x0}$,
both of which are regarded as effective models
for the present nonlocal model at large $g$ and $\lambda=0$ 
obtained by a simple
reduction of the dynamical degrees of freedom.
Two models are complementary each other and study of them
helps us to understand the properties of the  present nonlocal gauge model 
studied by the numerical simulations in section 4.
In particular, these models are capable to explain the existence of
the deconfinement (ordered) phase in the PD-1 model at large $g$ 
and the nonexistence of it in the other PD and ED models 
with $\lambda = 0$.


\subsection{The 1D XY model with spatial variables $U_{xi}$}

Let us imagine the situation
in which fluctuations of the temporal gauge fields $U_{x0} $ 
are small so that one may replace $U_{x0} = 
\exp(i\theta_{x0})$ by its average as follows;
\begin{eqnarray}
U_{x0} \rightarrow u\  (0 < u < 1).
\label{temporal}
\end{eqnarray}
This is expected for sufficiently large $g$. 
Furthermore we put $\lambda =0$, i.e., there 
are no direct interactions among
spatial gauge variables $U_{xi}=\exp(i\theta_{xi})$. 
Then the system is decoupled
to 1D subsystems defined at each spatial link $(x,x+i)$.
The  subsystem at $(x,x+i)$ is described by the U(1) angles 
$\theta_{j} (\equiv \theta_{xi})$ 
(we write $x_0\equiv j=1,\cdots, N_0$ and suppress the suffix $x_1,x_2,i$). 
Its energy
takes the form of a 1D XY spin model with nonlocal interactions
with couplings $c_\tau$. The partition functions of
the subsystem and the total system are given as follows;
\begin{eqnarray}
&& Z_{\rm 1DXY} =  \prod_{j=1}^{N_0} \int_{-\pi}^{\pi}
\frac{d\theta_{j}}{2\pi}\exp\left[g \sum_{j=1}^{N_0}\sum_{\tau=1}^{N_0}  
u^{2\tau} c_{\tau}\cos(\theta_{j+\tau}-\theta_{j})\right],\ 
\nonumber\\
&& Z_{\cal T}(g:{\rm large},\lambda=0) \simeq (Z_{\rm 1DXY})^{2N_1N_2}.
\label{ZZ}
\end{eqnarray} 
To study the correlation function 
$\langle \exp[i(\theta_{r}-\theta_{0})]\rangle$, 
we make the harmonic approximation for large $g$;
(i) expand the cosine term
up to the quadratic term, (ii) extend the range of $\theta_{j}$
to $(-\infty, \infty)$ and (iii) neglect topologically nontrivial
configurations. 
We expect that the third approximation in the above is justified
by the {\em long-range ferromagnetic} interactions in 
Eq.(\ref{ZZ}).
Then we have
\begin{eqnarray}
Z_{\rm 1DXY} &\simeq& 
\prod_{j=1}^{N_0} \int_{-\infty}^{\infty}
\frac{d\theta_{j}}{2\pi}\exp\left[ -\frac{g}{2}\sum_{j=1}^{N_0}
\sum_{\tau=1}^{N_0}  u^{2\tau}
c_{\tau}\left(\theta_{j+\tau}-\theta_{j}\right)^2\right]
\nonumber\\
&\propto& \prod_{k=1}^{N_0} \int_{-\infty}^{\infty}
d\tilde{\theta}_k \exp\left[ -\sum_{k=1}^{N_0}
G_k \tilde{\theta}_k^2\right],\nonumber\\
G_k&=& g \sum_{\tau} u^{2\tau} c_{\tau}[1-\cos(k\tau)], \nonumber\\
f(r) &\equiv& \langle \exp[i(\theta_{r}-\theta_{0})]\rangle
= \exp[-\frac{1}{2}\langle(\theta_{r}-\theta_{0})^2\rangle] \nonumber\\
&=& \exp[-\sum_k G^{-1}(k)(1-\cos(kr))],
\end{eqnarray} 
where $\tilde{\theta}_k$ is the Fourier-transformed variable of $\theta_{j}$.
For the standard local coupling $c_\tau = \delta_{\tau 1}$,
the $k$-sum in the exponent of $f(r)$ gives $\int dk \cos(kr)/k^2 \sim r$,
which implies $f(r) \simeq \exp[-(gu^2)^{-1}r] 
\rightarrow 0$ as $r \rightarrow
\infty$ due to the severe infrared fluctuations. 
For the nonlocal cases we have
\begin{eqnarray}
f(r) &\simeq& \exp\left[-\sum_k \frac{1-\cos(kr)}{g \sum_{\tau} 
u^{2\tau} c_{\tau}\tau^2 k^2}\right]
\simeq \exp(-M r), \nonumber\\
M &=& \frac{1}{g \sum_{\tau} u^{2\tau} c_{\tau}\tau^2 }\rightarrow
\left\{
\begin{array}{ll}
 0, & {\rm PD-1}\\
\frac{1-u^2}{gu^2}, & {\rm PD-2}   \\
-\frac{1}{g \ln(1-u^2)},& {\rm PD-3}\\
\frac{(1-h)^3}{gh(1-h+3h^2-h^3)},\
(h = u^2 e^{-1}) & {\rm ED}
\end{array}
\right..
\label{1dxy}
\end{eqnarray} 
Namely,  the order in the correlation function survives 
as $N_0 \rightarrow \infty$ for the  PD-1 case if $u\neq 0$, whereas 
the order is destroyed in the other cases even for $u\neq 0$. 
This result is just consistent with the numerical results 
for specific heat and the Polyakov lines studied
in section 4. 
For $\lambda=0$ case, the ordered-deconfinement phase 
is realized only in the PD-1 model at large $g$.  


\subsection{The 2D XY model with temporal variables $U_{x0}$}

The discussion in the previous subsection is supported 
by considering another effective model that focuses on the dynamics of 
the temporal gauge field $U_{x0}$.
To obtain it, we first replace the spatial gauge field $U_{xi}$ by its average,
\begin{eqnarray}
U_{xi} \rightarrow v \ \ (0 < v < 1).
\label{spatial}
\end{eqnarray}
This replacement is supported for the PD-1 model by the result
of the 1D XY model (\ref{1dxy}) in the previous subsection.
Then we fix the gauge to the temporal gauge by setting 
$U_{x0}= 1$ except for $x_0 =N_0$ because of the periodic boundary condition.
By keeping only the terms in the energy 
involving $U_{x_{\bot},N_0,0}$, we have
\begin{eqnarray}
U_{x_{\bot},N_0,0}&\equiv& \exp(i\varphi_{x_{\bot}}),\nonumber\\
Z_{\rm 2DXY} &=&  \prod_{x_{\bot}} \int_{-\pi}^{\pi}
\frac{d\varphi_{x_{\bot}}}{2\pi}\exp(A_{\rm 2DXY}),\nonumber\\
A_{\rm 2DXY} &=& gv^{2} \sum_{x_{\bot}}
\sum_{i=1,2}\sum_{\tau=1}^{N_0} c_{\tau}\cos(\varphi_{x_{\bot}+i}-
\varphi_{x_{\bot}}).
\end{eqnarray} 
This is just the standard 2D XY spin model having 
the {\em nearest-neighbor couplings} 
among the XY spins $\exp(i\varphi_{x_{\bot}})$. 
Here the XY spin correlation function corresponds to the 
correlation function of the Polyakov lines studied in section 4,
\begin{eqnarray}
\langle \cos(\varphi_{x_{\bot}}-\varphi_{0_{\bot}})\rangle = f_P(x_{\bot}).
\end{eqnarray}

The spin stiffness in the effective 2D XY model is given as
\begin{eqnarray}
g v^2 Q_1 \rightarrow
\left\{
\begin{array}{ll}
 \infty, & {\rm PD-1}\\
\frac{gv^2 \pi^2}{6}, & {\rm PD-2}   \\
\frac{gv^2 \pi^4}{90},& {\rm PD-3}\\
\frac{gv^2}{e-1}, & {\rm ED}
\end{array}
\right..
\label{2dxy}
\end{eqnarray}
It is well known that the 2D XY model with finite spin stiffness
has no long-range order, although the Kosterlitz-Thouless
transition is possible. Only in the PD-1 case, the spin stiffness diverges
for finite $v$
and the model may have the order $\langle U_{x_{\bot},N_0,0} \rangle \neq 0$.
This result (\ref{2dxy}) supports the procedure (\ref{temporal}) 
for the 1D XY model in the previous subsection for the PD-1 case.
On the other hand,  Eq.(\ref{1dxy})  
supports the replacement (\ref{spatial}) as we mentioned before.
Thus studies of these two XY models give us the
conclusion that only the PD-1 model at large $g$ has
the ordered (Coulomb) phase for the $\lambda=0$ case.

Studies of the two effective XY models in this section predict that 
the cases of PD-2, PD-3 and ED with $\lambda=0$ have no Coulomb phase,
as it was verified by the previous numerical calculations. 
Inclusion of $\lambda$ term, however, generates direct interactions
between the spatial variables $U_{xi}$. 
For small $\lambda$, $U_{xi}$'s ($i=1,2$) can be integrated out perturbatively
in powers of $\lambda$.
The obtained effective model contains nonlocal and multi-body
interactions of $\varphi_{x_{\bot}}$'s which prefer the ferromagnetic order
of $\varphi_{x_{\bot}}$.
Therefore the effective model for $\lambda \sim 1$ may {\em not} belong to
the same universality class of the 2DXY model with local interactions.
Then the above result does {\em not} contradict the numerical result in the
previous section, which shows the deconfinement phase exists in the PD-2
model with $\lambda=1$.
 

\section{Conclusion}
\setcounter{equation}{0}

In this paper, we studied the nonlocal compact U(1) gauge theory
 on the 3D lattice,
which ``simulates"  gauge models coupled with massless/massive
matter fields.
The main contributions of the present paper may be the following
two points: (i) MC simulations are feasible within reasonable
computer time even for 3D
lattice gauge theories with nonlocal interactions along one direction,
and (ii) the measurements of $E$,$C$, Polyakov lines, Wilson loops, 
and local and nonlocal instantons give rise to clear and  consistent results
on the phase structure of the model. In particular, they
distinguish the cases of ND($\lambda=1,0$), PD-1($\lambda=1,0$), 
PD-2($\lambda=1$) with a CDPT from the other cases of PD-2($\lambda=0$),
PD-3,ED without CDPT in a definitive manner.

As explained in Sect.2.3, the results obtained in this paper 
are quite important for studies of 
the strongly-correlated electron
systems like the high-$T_c$ cuprates, the fractional 
quantum Hall effect, quantum spin models, etc.
For example, in the t-J model of high-$T_c$ superconductivity, 
by using the hopping expansion of holons and spinons 
at finite $T$ with the continuous imaginary time, 
we derived an effective gauge
theory, which is highly nonlocal in the temporal direction.
The obtained effective theory has a similar action
as Eq.(\ref{simplified-model}) with $c_\tau=$constant and 
$g \propto n$ where $n$ is the density of matter 
fields(holons and spinons)\cite{pfs}.
This corresponds to the ND model.
Although the above effective gauge model is obtained for the system 
at finite $T$, we expect that a similar gauge model appears as an 
effective model at $T=0$. The result that the ND model has a CDPT
strongly suggests that the t-J model has the corresponding
phase transition into the deconfinement phase, which is nothing but
the charge-spin separated phase.

In the deconfinement phase, 
all the three variables $U_{x\mu}\ (\mu=0,1,2)$ are stable,
having small fluctuations.
The stability of the Lagrange multiplier $U_{x0}$ 
means that the constraint
(\ref{constraint}) is {\em not} respected by the holons and spinons,
the low-energy excitations of the system, whereas
the stability of $U_{xi}$ indicates that the gauge interaction
between the holons and spinons can be treated perturbatively.
Therefore, quasi-particles in the CSS state are the holons,
spinons and weakly interacting gauge bosons\cite{pfs}.
Of course, in order to present a definite ``proof" of the CSS,
it is necessary  to investigate a gauge system with full 
isotropic nonlocal interaction, because the integration
over holons and spinons generates nonlocal interactions  
not only in the temporal direction but also in the spatial
directions and their combinations.

Another interesting model related with the present one is the U(1) 
Higgs model
coupled with the nonlocal gauge field.
At present, 
it is believed that there is {\em no phase transition} in the 3D 
U(1) gauge-Higgs model with the ordinary local action if the Higgs field
has the fundamental charge and its radial fluctuations are 
suppressed\cite{AHM}.
However, the situation  may be changed by nonlocal gauge interactions. 
The existence of the deconfinement phase in the present nonlocal 
gauge system without Higgs fields
suggests that all the three phases, i.e., the confinement, 
Coulomb and Higgs phases,
may be realized in the 3D nonlocal gauge model with a 
local coupling to a Higgs field. The deconfinement phase
of the present model corresponds to the Coulomb phase.
This problem is closely 
related with ``doped holes" in the algebraic spin liquid
which may be realized in certain antiferromagnetic spin models
and materials in the spatial 2D lattice.
We shall report on these problems in a separate 
publication\cite{TSIM}. \\

{\bf Acknowledgement}

One of the authors (K.S.) thanks the members of 
Department of Physics, Kanazawa 
University for their hospitality delivered to him during his stay.

\newpage

\appendix 

\renewcommand{\theequation}{A.\arabic{equation}} 
\section{High- and Low-Temperature Expansions}
\setcounter{equation}{0}

In this appendix we study the behavior of the nonlocal gauge theory
(\ref{simplified-model})
in two regions of $g$ by analytic methods; 
the region of small $g (<<1)$ by high-temperature expansion 
(HTE) in Sect.A.1 and the region of large $g (>>1)$ by 
low-temperature expansion (LTE) in Sect.A.2.
Once one obtains an approximate expression
for the partition function $Z_{\cal T}$,
one can calculate various thermodynamic quantities
like $E$ and $C$ in Eq.(\ref{EC}).


\subsection{High-temperature expansion (HTE) for small $g$}

\begin{figure}[htbp]
\begin{center}
\leavevmode
\epsfxsize=6cm     
\epsffile{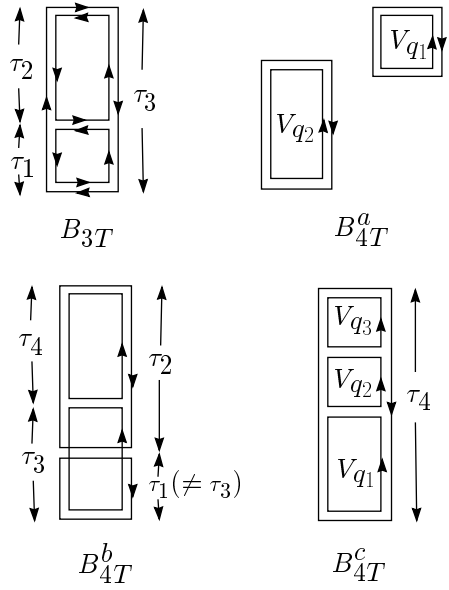}
  \caption{Illustration of the terms (\ref{htecoef}) in HTE.
  Each rectangle represents $V_{x,i,\tau}.$
  The terms $B_{3T},B_{4T}^{b,c}$ reflect the nonlocal nature;
they are absent in models of local
interaction characterized by $c_{\tau} = 0\ (\tau > 1)$. 
}
\label{fig-hte}
\end{center}
\end{figure}

Let us consider the case of small $g$. Since the action
$A_{\cal T}$ 
is proportional to $g$, one may expand the partition 
function $Z_{\cal T}$ of (\ref{simplified-model})
in powers of $g$ as
\begin{equation}
Z_{\cal T} = \int[dU] \exp(A_{\cal T}) =\int[dU]
\sum_{n=0}^{\infty}\frac{(A_{\cal T})^n}{n!}.\
\end{equation}
We obtain the expansion up to $O(g^4)$ as
\begin{equation}
Z_{\cal T} = 1+ \left(B_{2T}  + B_{2S} \lambda^2 \right) g^2
+ B_{3T} g^3+\left(B_{4T} + B_{4TS} \lambda^2 +B_{4S} 
\lambda^4 \right)g^4 +O(g^5),
\end{equation}
where each coefficient is expressed as 
\begin{eqnarray} 
B_{nT} &=& \frac{1}{n!}\int [dU] \left[ 
\sum_{q}c_{\tau}(V_q+ \bar{V}_q)\right]^n,\ 
q\equiv(x,i,\tau),\nonumber\\
B_{nS} &=& \frac{1}{n!}\int [dU] \left[ 
\sum_{p}(U_p+ \bar{U}_p)\right]^n, \;
U_p\equiv 
\bar{U}_{x+\hat{2},1}U_{x+\hat{1},2}U_{x1}\bar{U}_{x2},
\nonumber\\
B_{4TS}&=& \frac{{}_4C_2}{4!}\int [dU] \left[
\sum_{p}(U_p+ \bar{U}_p)\cdot
\sum_{q}c_{\tau}(V_q+ \bar{V}_q)\right]^2.
\end{eqnarray}
To evaluate the above coefficients, we use the following
formula for U(1) integral,
\begin{eqnarray}
\int dU_{x\mu} \bar{U}^m_{x\mu}U^{n}_{y\nu} = 
\delta_{xy}\delta_{\mu\nu}\delta_{mn}.
\end{eqnarray}
Then we obtain the following result;
\begin{eqnarray} 
B_{2T} &=& \sum_q c_\tau^2 = 2 V Q_2, \ Q_2 \equiv
\sum_{\tau} c_\tau^2,\nonumber\\
B_{2S} &=& \sum_p 1 = V,
\nonumber\\
B_{3T} &=& \frac{3\cdot 2}{3!}\sum_{x}
\sum_i
\sum_{\tau_1}\sum_{\tau_2}c_{\tau_1}c_{\tau_2}c_{\tau_3}
V_{x,i,\tau_1}V_{x+\tau_1\hat{0},i,\tau_2}\bar{V}_{x,i,\tau_3}
\delta_{\tau_3,\tau_1+\tau_2} + {\rm c.c.} 
\nonumber\\
&=& 4V\tilde{Q}_3,\ \tilde{Q}_3\equiv\sum_{\tau_1=1}^N\sum_{\tau_2=1}^N
\sum_{\tau_3=1}^N c_{\tau_1}c_{\tau_2}c_{\tau_3}
\delta_{\tau_3,\tau_1+\tau_2},
\nonumber\\
B_{4T}&=& B_{4T}^a+B_{4T}^b+B_{4T}^c,\nonumber\\
B_{4T}^a&=& \frac{1}{4!}\sum_{q_1,q_2,q_3,q_4}
\prod_{\ell=1}^4c_{\tau_\ell}\cdot
{}_4C_2\left[\delta_{q_1q_2}\delta_{q_3q_4}+\delta_{q_1q_3}\delta_{q_2q_4}
-\delta_{q_1q_2}\delta_{q_2q_3}\delta_{q_3q_4}\right]\nonumber\\
&=&2V^2Q_2^2-\frac{1}{2}VQ_4,\ Q_4 \equiv 
\sum_{\tau}c_\tau^4, \nonumber\\
B_{4T}^b&=&\frac{4\cdot 3\cdot 2}{4!}\sum_{x,i}
\sum_{\tau_1,\tau_2,\tau_3,\tau_4}
c_{\tau_1}c_{\tau_2}c_{\tau_3}c_{\tau_4}
(1-\delta_{\tau_1\tau_3})\delta_{\tau_4, \tau_1+\tau_2-\tau_3}
V_{x,i,\tau_1}V_{x+\tau_1\hat{0},i,\tau_2}\bar{V}_{x,i,\tau_3}
\bar{V}_{x+\tau_3 \hat{0},i,\tau_4}\nonumber\\
&=& 2V(\tilde{Q}_{4-} -Q_2^2),\ 
\tilde{Q}_{4-} \equiv \sum_{\tau_1,\tau_2,\tau_3,\tau_4}
c_{\tau_1}c_{\tau_1}c_{\tau_3}c_{\tau_4}\delta_{\tau_4, \tau_1+\tau_2-\tau_3},
\nonumber\\
B_{4T}^c&=&\frac{4!}{4!}\sum_{x,i}
\sum_{\tau_1,\tau_2,\tau_3,\tau_4}
c_{\tau_1}c_{\tau_2}c_{\tau_3}c_{\tau_4}
V_{x,i,\tau_1}V_{x+\tau_1\hat{0},i,\tau_2}V_{x+(\tau_1+\tau_2)\hat{0},
i,\tau_3}\bar{V}_{x,i,\tau_4}
\delta_{\tau_4, \tau_1+\tau_2+\tau_3}
+{\rm c.c.}\nonumber\\
&=& 4V\tilde{Q}_{4+},\ 
\tilde{Q}_{4+} \equiv \sum_{\tau_1,\tau_2,\tau_3,\tau_4}
c_{\tau_1}c_{\tau_1}c_{\tau_3}c_{\tau_4}\delta_{\tau_4, \tau_1+\tau_2+\tau_3},
\nonumber\\
B_{4TS}&=& \frac{1}{4} \sum_p 2 \cdot \sum_{q} 2 c_\tau^2
= 2V^2 Q_2,\nonumber\\
B_{4S}&=& \frac{1}{4!}\sum_{p_1,p_2,p_3,p_4}{}_4C_2\left[\delta_{p_1p_2}
\delta_{p_3p_4}+\delta_{p_1p_3}\delta_{p_2p_4}
-\delta_{p_1p_2}
\delta_{p_2p_3}\delta_{p_3p_4}\right]\nonumber\\
&=&\frac{1}{2}V^2 -\frac{1}{4}V.
\label{htecoef} 
\end{eqnarray}
Some of them are illustrated in Fig.\ref{fig-hte}.


Finally, we obtain the expression of the partition function and 
the free energy as follows,
\begin{eqnarray}
Z_{\cal T} &=& \exp(-FV),\nonumber\\
F&=&-(2Q_2+\lambda^2)g^2-4\tilde{Q}_3 g^3
+\left(\frac{Q_4}{2}+\frac{\lambda^4}{4} -4\tilde{Q}_{4+}-2\tilde{Q}_{4-}
\right)g^4 
+O(g^5).
\label{hte}
\end{eqnarray}
In the following, we list up the values of $Q_2$ and $Q_4$ 
for $N\rightarrow \infty$,
\begin{center}
\begin{eqnarray}
\begin{tabular}{|c|c|c|} 
\hline
       & $Q_2$     &  $Q_4$ 
\\ \hline
PD-1 &  $\frac{\pi^2}{6}=1.64493$ &  $\frac{\pi^4}{90}=1.08232$
\\ \hline
PD-2  &  $\frac{\pi^4}{90}$ & $\frac{\pi^8}{9450}=1.00408$
  \\ \hline
PD-3  & $\frac{\pi^6}{945}=1.01734$ & 1.00025   
\\ \hline
ED &  0.15652 &  0.018657 \\ \hline
\end{tabular}
\label{table}
\end{eqnarray}
\end{center}
By using  $Z_{\cal T}$ of 
Eq.(\ref{hte}), we obtain the internal energy $E$ and the specific
heat $C$ defined by Eq.(\ref{EC}) as
\begin{eqnarray}
E&=& -2(2Q_2+\lambda^2)g^2-12\tilde{Q}_3g^3
+4\left(
\frac{Q_4}{2}+\frac{\lambda^4}{4}-4\tilde{Q}_{4+}
-2\tilde{Q}_{4-}\right)g^4 + O(g^5),\nonumber\\
C&=&2(2Q_2+\lambda^2)g^2+24\tilde{Q}_3g^3
-12\left(\frac{Q_4}{2}+\frac{\lambda^4}{4}-4\tilde{Q}_{4+}
-2\tilde{Q}_{4-}\right)g^4 + O(g^5).
\label{hteec}
\end{eqnarray}


\begin{figure}[htbp]
\begin{center}
\leavevmode
\epsfxsize=5.5cm     
\epsffile{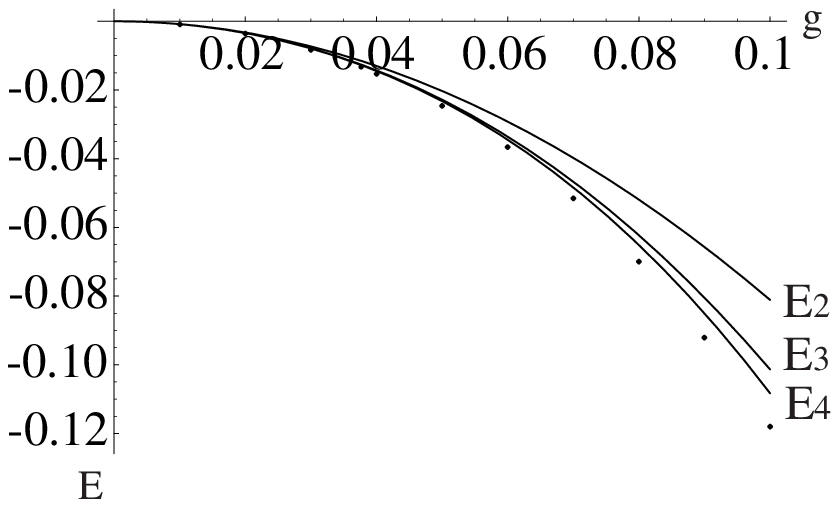}
\hspace{1cm}
\epsfxsize=5.5cm 
\epsffile{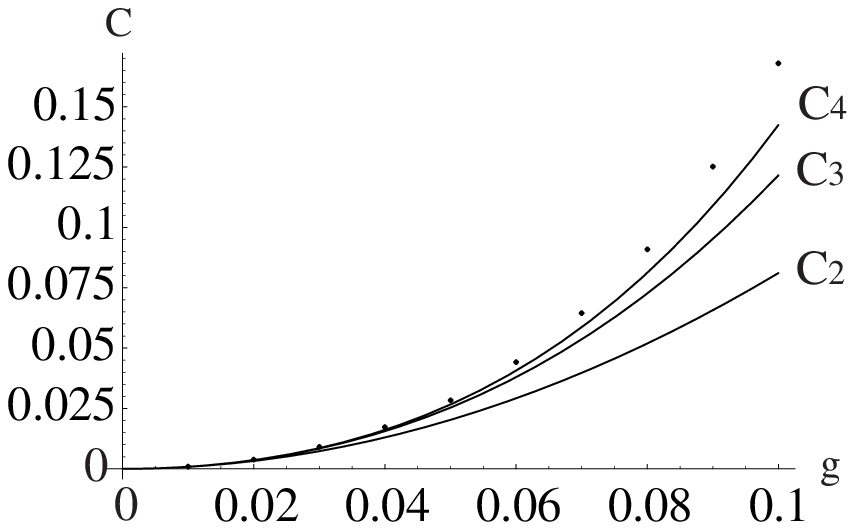}
  \caption{Comparison of HTE and the MC simulation.
We compare the HTE results and the MC results of $E$ and $C$ 
for the PD-1 model with $N=8$ and $\lambda=1$. The curves indicated by
$E_n$ and $C_n$ are the HTE results
with the terms up to $O(g^n)$. The dots are the MC data. 
}
\label{fig-hte-mc}
\end{center}
\end{figure}

\noindent
Fig.\ref{fig-hte-mc} shows that, 
as one includes the higher-order terms, the HTE results
approaches the MC result systematically. However, the 
approach is rather slow compared with the related models of local
interactions like the 3D XY spin model or the 3D U(1) pure LGT.
This is because the present nonlocal interactions
generate various important higher-order terms in the HTE that 
are absent from the local models.

Let us comment on the  convergence of the HTE. 
As usual, the HTE is an expansion in the disordered (confinement) phase
in which $U_{x\mu}$ fluctuates wildly.
Equation (\ref{hte}) shows that the convergence radius $g_{\rm HTE}$ 
of the expansion
is finite $g_{\rm HTE} \neq 0$, because  both the harmonic numbers
$Q_2$ and $Q_4$ appearing 
in the  coefficients are finite.
This means that
there exists certainly the finite region $0 \leq g^2 < g^2_{\rm HTE}$
of the confinement phase. We notice that if the long-range interaction
$c_\tau$ is very strong such that  $Q_2 = \infty$, the confinement phase 
may disappear.


\subsection{Low-Temperature Expansion (LTE) for large $g$}

For large $g$, we evaluate $Z_{\cal T}$ by the LTE.
The LTE is an expansion in powers of $g^{-1}$ 
around a fixed ``lowest-energy configuration"
of $U_{x\mu}$ like $U_{x\mu}=1$, which gives rise to 
the global maximum of $A_{\cal T}$.
Let us expand $U_{x\mu}$ as
\begin{eqnarray}
U_{x\mu}\equiv \exp(i\theta_{x\mu}) = 1+i\theta_{x\mu}-\frac{1}{2}
\theta_{x\mu}^2+ O(\theta_{x\mu}^3),
\label{exp}
\end{eqnarray}
where $\theta_{x\mu}$ is treated as $O(g^{-1/2})$
as we shall see. 
$A_{\cal T}$ is expanded up to the second order in 
$\theta_{x\mu}$ in the following quadratic form;
\begin{eqnarray}
A_{\cal T} = 4gQ_1V+2g\lambda V -g\sum_{x,\mu}\sum_{y,\nu}
\theta_{x\mu}G_{x\mu,y\nu}(\lambda)\theta_{y\nu} +O(\theta^4),
\end{eqnarray}
where the first two terms  $4gQ_1V+2g\lambda V$ 
come from the
first term, unity,  of R.H.S. of Eq.(\ref{exp}). 
Due to the gauge invariance, one may extend
the region of $\theta_{x\mu}$ from $\theta_{x\mu} \in (-\pi,\pi)$
to $\theta_{x\mu} \in (-\infty, \infty)$ together with a gauge
fixing. 
We take the temporal gauge $\theta_{x0}=0$ in the following calculation. 
Then we evaluate $Z_{\cal T}$ 
by rescaling $\theta_{xi}' = g^{1/2} \theta_{xi}$ and performing
Gaussian integration as
\begin{eqnarray}
Z_{\cal T} &\simeq&  e^{(4Q_1+2\lambda)gV}\prod_{x}\left[
 \int_{-\infty}^{\infty} d\theta_{x1}
\int_{-\infty}^{\infty} d\theta_{x2}\right]\exp\left(
-g\sum_{x,\mu}\sum_{y,\nu}
\theta_{x\mu}G_{x\mu,y\nu}\theta_{y\nu}\right)\nonumber\\
&=& 
e^{(4Q_1+2\lambda)gV}
\prod_{x}\left[ g^{-1} \int_{-\infty}^{\infty} d\theta'_{x1}
\int_{-\infty}^{\infty} d\theta'_{x2}\right]\exp\left(
-\sum_{x,\mu}\sum_{y,\nu}
\theta'_{x\mu}G_{x\mu,y\nu}\theta'_{y\nu}\right)\nonumber\\
&=&  \exp\left[(4Q_1g+2\lambda g -\ln g)V -\frac{1}{2}{\rm Tr}
\ln G(\lambda)\right],\nonumber\\
F&=&-(4Q_1+2\lambda) g +\ln g + O(g^0).
\end{eqnarray}
This gives 
\begin{eqnarray}
E&=& -(4Q_1+2\lambda) g+1+O(g^{-1}),\nonumber\\
C&=&1+O(g^{-1}).
\label{lte}
\end{eqnarray}
The higher-order terms
in $F$ are 
$O(g^{-n})\ (n \geq 0)$ which may be calculated by the usual 
perturbation theory.  





\newpage

\begin{thebibliography}{1}



\bibitem{im1} 
G.Baskaran and P.W.Anderson, Phys.Rev.B37(1988)580;\\
 A.Nakamura and T.Matsui, Phys.Rev.B37(1988)7940; \\
 D.P. Arovas and A. Auerbach,
Phys.Rev.B38(1988)316; \\
L.B.Ioffe and A.L.Larkin, Phys.Rev.B39(1989)8988; \\
I.Ichinose and T.Matsui, Phys.Rev.B45(1992)9976.

\bibitem{FQHE}R.B.Laughlin, Phys.Rev.Lett.50(1983)1395;\\
J.K.Jain, Phy.Rev.Lett.63(1989)199.

\bibitem{CSS}P.W.Anderson, Phy.Rev.Lett.64(1990)1839.

\bibitem{pfs}
For the quantum Hall states, see 
I.Ichinose and T.Matsui, Phys.Rev.B68(2003)085322
and the references cited therein.
For the high-temperature  superconductivity, see 
I.Ichinose and T.Matsui, 
Nucl.Phys.B394(1993)281;
Phys.Rev. B51(1995)11860;
I.Ichinose and T.Matsui, and M.Onoda, Phys.Rev.B64(2001)104516.  

\bibitem{nayak}
C.Nayak, Phys.Rev.Lett.85, 178(2000);
I.Ichinose and T.Matsui, Phys.Rev.Lett.86, 942(2001).
See also  Sect.VI of the last reference of Ref.\cite{pfs}.

\bibitem{FS}E.Fradkin and S.H.Shenker, Phys.Rev.D19(1979)3682.

\bibitem{QCD}Y.Iwasaki, K.Kanaya, S.Sakai, and T.Yoshie,
Phys.Rev.Lett.69(1992)21.

\bibitem{QED}J.B.Kogut and C.G.Strouthos,
Phys.Rev.D67, 034504(2003) 
and references cited therein; \\
S.Hands and I.O.Thomas, Phys.Rev.B72(2005)054526; \\
G.Grignani, G.Semenoff and P.Sodano, 
Phys.Rev.D53(1996)7157.

\bibitem{AFH}
The appearance of QED$_3$ is argued explicitly  
for the antiferromagnetic Heisenberg model in
I. Affleck and J.B. Marston, Phys.Rev.B37(1988)3774, and 
for the slave-fermion t-J model in 
the last reference of Ref.\cite{im1}.

\bibitem{AIMS}A part of the results obtained in this paper has already 
been reported in G.Arakawa, I.Ichinose, T.Matsui, and K.Sakakibara, 
Phys.Rev.Lett.94 (2005) 211601.

\bibitem{polyakov}A.M.Polyakov, Nucl.Phys.B120(1977)429.

\bibitem{QED3}
H.Kleinert, F.S.Nogueira, and A.Sudb\o, Phys.Rev.Lett.88(2002)232001; \\
Nucl.Phys.B666(2003)361; \\
F.S.Nogueira and H.Kleinert, cond-mat/0501022.

\bibitem{css-}
I.F.Herbut and B.H.Seradjeh, Phys.Rev.Lett.91(2003)171601;\\
I.F.Herbut, B.H.Seradjeh, S.Sachdev, and G.Murthy, 
Phys.Rev.B68(2003)195110.


\bibitem{plasma}
S. Kragset, A. Sudb\o, and F. S. Nogueira,
Phys. Rev. Lett. 92 (2004)186403;\\
K. B\o rkje, S. Kragset, and A. Sudb\o,
Phys. Rev. B 71 (2005) 085112.

\bibitem{CPN}I.Ya.Aref'eva and S.I.Azakov, Nucl.Phys.B162(1980)298;
A.D'Adda, P.Di Vecchia and M. L\"uscher,
Nucl.Phys.B146, 63(1978); E.Witten, Nucl.Phys.B149, 285(1979).

\bibitem{yoshioka}
See, e.g., S.Sachdev, 
{\it ``Quantum Phase Transitions"}, (Cambridge University Press,
Cambridge, England, 1999);\\
D.Yoshioka, G.Arakawa, I.Ichinose and T.Matsui, 
Phys.Rev. B70, 174407(2004); \\
B.A.Bernevig, D.Giuliano, and R.B.Laughlin, Annals Phys.311(2004)182;\\
Ki-Seok Kim, cond-mat/0406511.

\bibitem{TIM}S.Takashima, I.Ichinose, and T.Matsui, 
Phy.Rev.B72(2005)075112, cond-mat/0504193.

\bibitem{CST}See for example, G.W.Semenoff, P.Sodano and Y.S. Wu,
Phys.Rev.Lett.62(1989)715, and references cited therein.

\bibitem{IN}See for example, I.Ichinose and M.Onoda, 
Nucl.Phys.B435[FS](1995)637
and references cited therein.

\bibitem{FSS} See, for example, J.M.Thijssen, {\it ``Computational
Physics"} (Cambridge University Press, 1999).

\bibitem{instanton}T.A.DeGrand and D.Toussaint,
Phys.Rev.D22(1980)2478. 

\bibitem{crossover}R.J.Wensley and J.D.Stack, Phys.Rev.Lett.63(1989)1764.

\bibitem{AHM}
See, for example, M.N.Chernodub, E.-M.Ilgenfritz and A.Schiller,
Phys.Lett.B547(2002)269;
S.Wenzel, E.Bittner, W.Janke, A.M.J.Schakel, and A.Schiller,
Rhys.Rev.Lett.95(2005)051601.

\bibitem{TSIM}S.Takashima, K.Sakakibara, I.Ichinose, and T.Matsui,
paper in preparation.

\end{thebibliography}
\end{document}